\documentclass[preprint,prd,nofootinbib,tightenlines,amsmath]{revtex4}
\usepackage{enumerate}
\usepackage{multirow}
\usepackage[utf8]{inputenc}
\usepackage{anysize}
\usepackage{textcomp}
\usepackage{epsfig}
\usepackage{graphics,color} 
\usepackage{verbatim}
\usepackage{afterpage}
\usepackage{amsmath}    
\usepackage{soul}
\usepackage{xcolor,cancel}

\usepackage{blindtext}

\catcode`\@=11
\def\sla@#1#2#3#4#5{{%
 \setbox\z@\hbox{$\m@th#4#5$}%
 \setbox\tw@\hbox{$\m@th#4#1$}%
 \dimen4\wd\ifdim\wd\z@<\wd\tw@\tw@\else\z@\fi
 \dimen@\ht\tw@
 \advance\dimen@-\dp\tw@ \advance\dimen@-\ht\z@
 \advance\dimen@\dp\z@
 \divide\dimen@\tw@ \advance\dimen@-#3\ht\tw@
 \advance\dimen@-#3\dp\tw@ \dimen@ii#2\wd\z@
 \raise-\dimen@\hbox to\dimen4{%
 \hss\kern\dimen@ii\box\tw@\kern-\dimen@ii\hss}%
 \llap{\hbox to\dimen4{\hss\box\z@\hss}}}}

\def\slashed#1{%
 \expandafter\ifx\csname sla@\string#1\endcsname\relax
{\mathpalette{\sla@/00}{#1}}
\fi}
\def\declareslashed#1#2#3#4#5{%
 \expandafter\def\csname sla@\string#5\endcsname{%
#1{\mathpalette{\sla@{#2}{#3}{#4}}{#5}}}}
 \catcode`\@=12
\declareslashed{}{/}{.08}{0}{D}
 \declareslashed{}{/}{.1}{0}{A}
 \declareslashed{}{/}{0}{-.05}{k}
 \declareslashed{}{/}{.1}{0}{\partial}
 \declareslashed{}{\not}{-.6}{0}{f}

\def\lsim{\mathrel {\vcenter {\baselineskip 0pt \kern 0pt
    \hbox{$<$} \kern 0pt \hbox{$\sim$} }}}
\def\gsim{\mathrel {\vcenter {\baselineskip 0pt \kern 0pt
    \hbox{$>$} \kern 0pt \hbox{$\sim$} }}}

\newcommand{\bea}{\begin{eqnarray}}
\newcommand{\eea}{\end{eqnarray}}

\begin{document}

\baselineskip=15pt
\preprint{}

\title{Spin correlations and new physics in $\tau$-lepton decays at the LHC}

\author{Alper Hayreter$^{1}$\footnote{Electronic address: alper.hayreter@ozyegin.edu.tr} and German Valencia$^{2}$\footnote{Electronic address: valencia@iastate.edu }}


\affiliation{$^{1}$ Department of Natural and Mathematical Sciences, Ozyegin University, 34794 Istanbul Turkey.}

\affiliation{$^{2}$ Department of Physics, Iowa State University, Ames, IA 50011.}

\date{\today}

\vskip 1cm
\begin{abstract}

We use spin correlations to constrain anomalous $\tau$-lepton couplings at the LHC including its anomalous magnetic moment, electric dipole moment and weak dipole moments. Single spin correlations are ideal to probe interference terms between the SM and new dipole-type couplings as they are not suppressed by the $\tau$-lepton mass. Double spin asymmetries give rise to $T$-odd correlations useful to probe  $CP$ violation purely within the new physics amplitudes, as their appearance from interference with the SM is suppressed by $m_\tau$. We compare our constraints to those obtained earlier on the basis of deviations from the Drell-Yan cross-section.

\end{abstract}

\pacs{PACS numbers: }

\maketitle

\section{Introduction}

There exists a very active $\tau$-lepton program at the ATLAS and CMS collaborations. They have already studied the reconstruction of the $Z$-resonance in the di-tau mode \cite{Chatrchyan:2011nv,Aad:2011kt} and placed limits on physics  beyond the standard model (BSM). Searches for heavy resonances such as $Z^\prime$ bosons have excluded such particles in the di-tau channel up to masses around 1~TeV \cite{Chatrchyan:2012hd,Aad:2012gm,Aad:2015osa}. There is also an active program to observe the Higgs boson in this channel \cite{Aad:2012mea,Chatrchyan:2012vp,Chatrchyan:2014nva,Aad:2015vsa}, as well as searches for additional neutral scalars that decay to $\tau^+\tau^-$ \cite{Khachatryan:2014wca}. In this paper we propose the use of spin correlations to constrain $\tau$ anomalous couplings.

An effective Lagrangian for BSM physics with a complete catalog of operators up to dimension six exists in the literature \cite{Buchmuller:1985jz,Grzadkowski:2010es} for the case in which the observed 126~GeV state is the Higgs boson in the SM. In a recent paper \cite{Hayreter:2013vna} we studied the subset of dimension six operators that describe the $\tau$-lepton dipole-type couplings, as well as two dimension eight operators which couple tau leptons directly to gluons and are thus enhanced by the gluon luminosities \cite{Potter:2012yv}. In particular we discussed the constraints that can be imposed on these operators by studying deviations from the Drell-Yan cross section at the LHC as well as by bounding the cross section for production of $\tau$-leptons in association with a Higgs boson. We now extend that study to include spin correlations measured in the angular distributions of muons or electrons in leptonic decay modes.

The couplings involved in the study are the $\tau$-lepton anomalous magnetic moment and electric dipole moment given by $a_\tau^\gamma$ and $d_\tau^\gamma$ respectively,
\begin{eqnarray}
{\cal L}=\frac{e}{2}\ \bar{\tau}\ \sigma^{\mu\nu}\left(a_\tau^\gamma+i\gamma_5 d_\tau^\gamma \right) \ \tau \ F_{\mu\nu}
\label{defedm}
\end{eqnarray}
and its corresponding weak dipole moments $a_\tau^Z$ and $d_\tau^Z$,
\begin{eqnarray}
{\cal L}=\frac{g}{2\cos\theta_W}\ \bar{\tau}\ \sigma^{\mu\nu}\left(a_\tau^Z+i\gamma_5 d_\tau^Z \right) \ \tau \ Z_{\mu\nu}
\label{defzdm}
\end{eqnarray}
The $\tau$-lepton dipole moments have been studied many times before in the literature \cite{Donoghue:1977bw, Silverman:1982ft, Barr:1988mc,delAguila:1990jg,Goozovat:1991nu,delAguila:1991rm,Bernreuther:1993nd,Cornet:1995pw,Vidal:1998jc,GonzalezSprinberg:2000mk,Bernabeu:2004ww,Bernabeu:2008ii}.

These anomalous  couplings, Eq.~\ref{defedm}~and~Eq.~\ref{defzdm}, originate in the gauge invariant dimension six operators, in the notation of \cite{Buchmuller:1985jz},
\begin{eqnarray}
{\cal L} = g\frac{d_{\ell W}}{\Lambda^2}\ \bar{\ell}\sigma^{\mu\nu}\tau^i e\  \phi W^i_{\mu\nu} + g^\prime\frac{d_{\ell B}}{\Lambda^2}\ \bar{\ell}\sigma^{\mu\nu}e  \ \phi B_{\mu\nu} \  +\ {\rm h.c.}
\label{ginvedm}
\end{eqnarray}
The correspondence between 
these gauge invariant operators and the anomalous magnetic moment,  electric dipole moment (EDM) and weak dipole moment (ZEDM) of the leptons is given by\footnote{There is a typo in Eq. 4 of Ref.~\cite{Hayreter:2013vna} that propagates to the conversion of our bounds from the gauge invariant basis to the tau anomalous couplings corrected in an errata. We also use here a different, more convenient, normalization for $d_{\ell W}$.}
\begin{eqnarray}
a_\ell^\gamma = \frac{\sqrt{2}\ v }{\Lambda^2}{\rm Re}\left(d_{\ell B}-d_{\ell W}\right) &,& a_\ell^Z = -\frac{\sqrt{2}\ v }{ \Lambda^2}{\rm Re}\left(d_{\ell W}+\sin^2\theta_W\ (d_{\ell B}-d_{\ell W})\right)
\nonumber \\
d_\ell^\gamma = \frac{\sqrt{2}\ v }{\Lambda^2}{\rm Im}\left(d_{\ell B}-d_{\ell W}\right) &,&
d_\ell^Z = -\frac{\sqrt{2}\ v }{ \Lambda^2}{\rm Im}\left(d_{\ell W}+\sin^2\theta_W\ (d_{\ell B}-d_{\ell W})\right)
\label{convertunits}
\end{eqnarray}
where $v\sim 246$~GeV is the Higgs vacuum expectation value, $\theta_W$ is the usual weak mixing angle and $\Lambda$ is the scale of new physics, which we take as 1~TeV for our numerical study.

As argued in \cite{Potter:2012yv} the usual power counting for new physics operators is altered for dimension eight operators that couple a lepton pair directly to gluons due to the larger parton luminosities. This motivates our inclusion of the  ``lepton-gluonic" couplings for the $\tau$ in this study
\begin{eqnarray}
{\cal L} = \frac{g_s^2}{\Lambda^4}\left(d_{\tau G} \ G^{A\mu\nu}G^A_{\mu\nu} \bar \ell_L \ell_R \phi  +d_{\tau \tilde{G}} \ G^{A\mu\nu} \tilde G^A_{\mu\nu} \bar \ell_L \ell_R \phi \right)\ + {\rm h.~c.}
\label{taugluon}
\end{eqnarray}
Here $G^A_{\mu\nu}$ is the gluon field strength tensor and $\tilde G^{A\mu\nu} = (1/2)\epsilon^{\mu\nu\alpha\beta}G^A_{\alpha\beta}$ its dual. 
If we allow for $CP$ violating phases in the coefficients, $d_{\tau G}$ and $d_{\tau \tilde{G}}$, the resulting gluon-lepton couplings take the form
\begin{eqnarray}
{\cal L} &=& \frac{v}{\sqrt{2}}\frac{ g_s^2}{\Lambda^4}\left({\rm Re}(d_{\tau G}) \ G^{A\mu\nu}G^A_{\mu\nu}  +{\rm Re}(d_{\tau \tilde{G}}) \ G^{A\mu\nu} \tilde G^A_{\mu\nu}\right) \bar \ell \ell \nonumber \\
&+& i\frac{v}{\sqrt{2}}  \frac{ g_s^2}{\Lambda^4}\left({\rm Im}(d_{\tau G}) \ G^{A\mu\nu}G^A_{\mu\nu}  +{\rm Im}(d_{\tau \tilde{G}}) \ G^{A\mu\nu} \tilde G^A_{\mu\nu}\right) \bar \ell \gamma_5 \ell
\label{couplings}
\end{eqnarray}

In Table~\ref{tau:results} we summarize the 1$\sigma$ constraints that we obtained on the $\tau$-lepton anomalous magnetic moment, electric dipole moment and weak dipole moments assuming a 14\% measurement of the Drell-Yan cross-section at LHC14 in Ref.~ \cite{Hayreter:2013vna}. We compare them to the best existing constraints from Delphi \cite{Abdallah:2003xd}, Belle \cite{Inami:2002ah} and Aleph \cite{Heister:2002ik}. The results can be interpreted as a sensitivity to a NP scale $\Lambda \sim 0.5$~TeV. For comparison, the same measurement of the Drell-Yan cross-section constrains the NP scale of the dimension 8 gluonic couplings  $\Lambda \sim 1$~TeV.
\begin{table}[htb]
\begin{center}
\begin{tabular}{|c|c|c|c|c|}
\hline
& $m_\tau a_\tau^V$ LHC-14 & $m_\tau a_\tau^V$ existing & $m_\tau d_\tau^V$ LHC-14 & $m_\tau d_\tau^V$ existing \\ \hline
$V=\gamma$ & (-0.0054,0.0060) & (-0.026,0.007) Delphi & (-0.0057,0.0057) &
(-0.002,0.0041) Belle \\ \hline
$V=Z$ & (-0.0018,0.0020) & (-0.0016,0.0016) Aleph & (-0.0017,0.0017) &
(-0.00067,0.00067) Aleph \\ \hline
\end{tabular}
\end{center}
\caption{Summary of constraints for 1$\sigma$ bounds that can be placed on the $\tau$-lepton anomalous magnetic moment, electric dipole moment and weak dipole moments at LHC14 from Ref.~ \cite{Hayreter:2013vna} compared to existing bounds.}
\label{tau:results}
\end{table}
The cross-sections we used to obtain the constraints in Table~\ref{tau:results}, are approximately quadratic in the anomalous couplings indicating that the interference with the SM is very small. This is, of course, due to the fact that the interference between the SM and the dipole-type couplings  is suppressed by the $\tau$-mass.  

In this paper we extend our previous study considering constraints that arise from spin correlations. These spin correlations evade the helicity suppression of the interference terms in the cross-section and produce observables linear in the new physics couplings. In this way it is possible to improve  the constraints on the electric dipole moments and  to study their $CP$ violating nature through $T$-odd asymmetries.

\section{Spin correlations}

Spin correlations in $\tau$-pair production including anomalous dipole type couplings have been studied in Ref.~\cite{Bernreuther:1993nd}. In that paper, the spin density matrix for production of $\tau$-pairs in $e^+e^-$ colliders in the CM frame was constructed and combined with the decay matrix for polarized $\tau$ in its rest frame. That formalism exhibits the spin correlations explicitly but is not suited for our calculation. We want to construct (Lorentz scalar) correlations in terms of observable momenta at the LHC, namely,  the muon (or electron) momenta and the beam momentum. Furthermore, we want to measure the correlations with event simulations using  {\tt MadGraph5} \cite{MadGraph}. The main advantage of this approach is the ease in introducing different types of new physics with the aid of {\tt FeynRules} \cite{Christensen:2008py}. In this paper we limit ourselves to dilepton decays of the $\tau$ pairs, but in a future publication we will address the hadronic decay modes.

\subsection{$CP$ violating couplings}

The imaginary part of the effective couplings gives rise to electric and weak dipole moments of the $\tau$-lepton. These dipole moments are known to produce a double spin correlation linear in the anomalous coupling, of the form
\begin{equation}
{\cal O}_{2s}\sim m_\tau d_\tau^{Z,\gamma} \epsilon_{\mu,\nu,\alpha,\beta}\ p_{\tau^+}^\mu p_{\tau^-}^\nu s_{\tau^+}^\alpha s_{\tau^-}^\beta
\label{2scorr}
\end{equation}
This correlation originates in the interference between the $CP$ violating edm amplitude  and the $CP$ conserving SM amplitude.  In this case, however, the interference requires a fermion helicity flip and is therefore proportional to the $\tau$-lepton mass, resulting in a large suppression at the LHC. On the other hand, contributions that are quadratic in the new physics couplings do not suffer from this suppression, and Eq.~\ref{2scorr} is useful to probe terms of the form $\sim {\rm Re}(d_{\tau V})\ {\rm Im}(d_{\tau V})$, that is,
\begin{eqnarray}
{\cal O}_{2s}&\sim& d^{Z,\gamma}_\tau\  a^{Z,\gamma}_\tau\ \epsilon_{\mu,\nu,\alpha,\beta}\ p_{\tau^+}^\mu p_{\tau^-}^\nu s_{\tau^+}^\alpha s_{\tau^-}^\beta
\end{eqnarray}
and similar terms proportional to both the real and imaginary parts of the couplings $d_{\tau G}$ and $d_{\tau \tilde{G}}$.

With the muons (or electrons) in leptonic tau decay acting as spin analyzers this is measurable as
\begin{eqnarray}
{\cal O}_{ss} &=& \epsilon_{\mu,\nu,\alpha,\beta}\ p_{\tau^+}^\mu p_{\tau^-}^\nu p_{\mu^+}^\alpha p_{\mu^-}^\beta
\end{eqnarray}
which requires at least partial reconstruction of one $\tau$-momentum direction and may be better suited for hadronic decay channels.

To probe the anomalous couplings that violate $CP$ with terms in the differential cross-section that are linear in ${\rm Im}(d_{\tau V})$, but not proportional to the $\tau$ mass, we resort to single spin correlations. For example, for the parton level process $q\bar{q} \to \tau^+\tau^-$ one finds that the $Z$ exchange diagram leads to the $CP$-odd correlation
\begin{equation}
{\cal O}_{1s} \sim d_\tau^Z g_A (\hat{t}-\hat{u})\ \epsilon_{\mu,\nu,\alpha,\beta}(p_{1}- p_{2})^\mu p_{\tau^+}^\nu p_{\tau^-}^\alpha \left(s_{\tau^-}-s_{\tau^+}\right)^\beta
\label{O1seq}
\end{equation}
where $\hat{t}$, $\hat{u}$ and $\hat{s}$ are the parton level Mandelstam variables, $g_A$ is the axial vector coupling of the $Z$ to the charged leptons and 
we have neglected the smaller vector coupling, $g_V$. Note that the parton momenta $p_{1,2}$ appear in a symmetric combination (from the two antisymmetric factors, $(\hat{t}-\hat{u})$, and the explicit $(p_1-p_2)$) and therefore this correlation does not vanish after the symmetrization of $p_{1,2}$ that follows from the convolution with the parton distribution functions for the LHC $pp$ initial  state. 

In order to write $T$-odd correlations that are sensitive to Eq.~\ref{O1seq} and are expressed only in terms of observable momenta we note that:
\begin{itemize}
\item In leptonic $\tau$ decay, the spin is analyzed by the muon (or the electron) momentum. The simplest way to compute this is using the method of Ref.~\cite{Antipin:2008zx}, which shows that for leptonic  $\tau$ decay, Eq.~\ref{O1seq} becomes
\begin{eqnarray}
{\cal O}_{1s}^{\ell\ell} \sim (\hat{t}-\hat{u})\left( p_{\tau^-}\cdot  p_{\mu^-} \epsilon_{\mu,\nu,\alpha,\beta}p_{1}^\mu p_{2}^\nu p_{\tau^-}^\alpha p_{\mu^+}^\beta
+ p_{\tau^+}\cdot  p_{\mu^+} \epsilon_{\mu,\nu,\alpha,\beta}p_{1}^\mu p_{2}^\nu p_{\mu^-}^\alpha p_{\tau^+}^\beta\right)
\label{O1lepeq}
\end{eqnarray}

\item In the lab frame at the LHC the $\tau$-leptons are highly boosted so their three-momenta are very close to that of the muons. Further, in leptonic $\tau$ decay it is not possible to reconstruct the $\tau$ momentum completely. We then replace the $\tau$ momenta with the corresponding muon momenta in the lab frame obtaining
\begin{eqnarray}
{\cal O}_{1s}^{\ell\ell}\xrightarrow[]{lab} \propto (\hat{t}-\hat{u}) \epsilon_{\mu,\nu,\alpha,\beta}p_{1}^\mu p_{2}^\nu p_{\mu^-}^\alpha p_{\mu^+}^\beta
\label{O1mmeq}
\end{eqnarray}

\item In the lab frame, the sum and difference of the proton momenta are just the center of mass energy and the beam direction, and the two parton momenta appearing in Eq.~\ref{O1mmeq} have to be expressed in terms of these:
\begin{eqnarray}
P_1=\frac{\sqrt{S}}{2}(1,0,0,1) &,&
P_2=\frac{\sqrt{S}}{2}(1,0,0,-1)\nonumber \\
P \equiv P_1+P_2\ =\ \sqrt{S}(1,0,0,0) &,& 
q_{\rm beam} \equiv P_1-P_2\ =\ \sqrt{S}(0,0,0,1)
\end{eqnarray}

\item This leaves us with two possibilities:
\begin{eqnarray}
{\cal O}_1 &=&\left[ \vec{q}_{beam}\cdot(\vec{p}_{\mu^+}- \vec{p}_{\mu^-})\ \vec{q}_{beam}\cdot \left( \vec{p}_{\mu^+}\times \vec{p}_{\mu^-}\right)\right]_{lab}\nonumber \\
{\cal O}_2 &=&\left[ \vec{q}_{beam}\cdot(\vec{p}_{\mu^+}+\vec{p}_{\mu^-})\ \vec{q}_{beam}\cdot \left( \vec{p}_{\mu^+}\times \vec{p}_{\mu^-}\right)\right]_{lab}
\end{eqnarray}

\end{itemize}

The $CP$ properties of these two forms merit discussion. Since these correlations involve the beam momentum, a property of the initial state, they do not have definite transformation properties under $CP$. In fact, $CP$ transforms  LHC correlations into anti-LHC ($\bar{p}\bar{p}$ collider) ones.
If we consider the same correlations for a $p\bar{p}$ collider instead, we see that in this case the first one is $CP$-odd and can only be produced by the electric dipole moments. The second one, however, is $CP$-even and cannot be produced by electric dipole moments at a $p\bar{p}$ collider. This is a novel feature for colliders,  that does not occur at the parton level  where the pre-factor $q\cdot (p_{\tau^+}+p_{\tau^-})= (p_1-p_2)\cdot (p_1+p_2)$ vanishes. We illustrate this with an example  in Table~\ref{colliders}.

Noting again that the symmetry of the initial $pp$ state at the LHC forbids terms linear in $q_{\rm beam}$, we use the correlation
\begin{eqnarray}
{\cal O}_{test} &=&\left[ \vec{q}_{beam}\cdot \left( p_{\mu^+}\times p_{\mu^-}\right)\right]_{lab} 
\end{eqnarray}
to gauge the statistical significance of our asymmetries. Interestingly, as seen in Table~\ref{colliders}, this asymmetry does not vanish in $p\bar{p}$ colliders and would in fact be the most sensitive one to use in that case.

\subsection{$CP$ conserving couplings}

The $CP$ conserving dipole couplings interfere with the SM but this contribution  to the cross-section is suppressed by the $\tau$ lepton mass as well. It is also possible to find terms that are linear in the $CP$ conserving anomalous couplings and that are not helicity suppressed by looking at single spin correlations. One such term is given by
\begin{equation}
{\cal O}_{1spin} \sim a_\tau^Z\ g_A\ \left( \hat{s}(\hat{t}-\hat{u})\ (p_1-p_2)\cdot(s_{\tau^-}+s_{\tau^+})\ +\ (\hat{t}-\hat{u})^2\ (p_{\tau^+}s_{\tau^-}-p_{\tau^-}s_{\tau^+})\right)
\label{cpeven}
\end{equation}
To study Eq.~\ref{cpeven} using only the beam and muon momenta we found the following two observables: the first one is the muon charge asymmetry \cite{Gupta:2011vt} defined by
\begin{eqnarray}
{\cal O}_{C}&=& \Delta |y| \equiv |y_{\mu^+}|- |y_{\mu^-}|
\label{ac}
\end{eqnarray}
The charge asymmetry is $C$-odd and therefore changes sign at $\bar{p}\bar{p}$ collider and vanishes at a $p\bar{p}$ collider as seen in the example in Table~\ref{colliders}.
The second possibility is simply 
\begin{eqnarray}
{\cal O}_{p_T} &=& \vec{q}_{beam}\cdot( \vec{p}_{\mu^+}- \vec{p}_{\mu^-})\ \vec{q}_{beam}\cdot( \vec{p}_{\mu^+}+ \vec{p}_{\mu^-})
\label{o4}
\end{eqnarray}
which can also be written as the difference in transverse momentum of the two muons. 

To measure any of the correlations discussed above we use the fully integrated counting  asymmetries normalized to the standard model cross-section,
\begin{eqnarray}
A_i &=& \left(\frac{N_+ - N_-}{N_+ + N_-}\right)\left(\frac{\sigma}{\sigma_{SM}}\right)
\end{eqnarray}
where $N_{+}=\sigma({\cal O}_i > 0)$, and $N_{-}=\sigma({\cal O}_i < 0)$.
The normalization to the standard model cross-section, instead of the total cross-section including new physics, is chosen because it shows clearly whether a given asymmetry is linear or quadratic in the new couplings. Of course, if there is new physics large enough to be detected from deviations in the cross-section from its standard model value it is possible to simply scale the asymmetries we calculate here. It is also possible to look closer at details of the angular distributions to attempt to extract new physics, but here we limit ourselves to the overall counting asymmetries.

\section{Numerical study}

For our numerical study we generate multiple event samples for the process $pp \to \tau^+\tau^- \to \ell^+\ell^- \nu_\tau \bar\nu_\tau \nu_\mu \bar\nu_\mu$ (where $\ell =\mu,e$ but will be a muon for most of our study) at 14~TeV center of mass energy that we summarize in the Appendix.  The anomalous couplings are implemented in  {\tt MadGraph5} \cite{MadGraph} with the aid of {\tt FeynRules} \cite{Christensen:2008py} \footnote{The code is available from the authors upon request.}. We use the resulting UFO model files to generate events for several values of $d_{\tau W}$, $d_{\tau B}$, $d_{\tau G}$ and $d_{\tau \tilde{G}}$ in a range motivated by our previous results from Ref.~\cite{Hayreter:2013vna}. 
The events preserve all spin correlations between production and decay of the $\tau$-leptons as they are generated for the complete process. In each case we generate event samples with one million dimuon or dilepton events {\it after} cuts, implying a $1\sigma$ statistical sensitivity to all asymmetries at the $(\sigma/\sigma_{SM}\times 0.1)\%$ level. 

\subsection{High energy dilepton pairs}
The cuts used in our event generation are:
\begin{itemize}
\item $m_{\tau\tau} > 120$~GeV  implemented in the cuts.f file. The purpose of this cut is to exclude the $Z$ resonance region from consideration, as this will be discussed separately. We use this idealized cut for simplicity although it may not be possible to implement experimentally for leptonic tau decays. In a more realistic simulation removal of the $Z$ region can be effectively accomplished with an alternative cut on $m_{\mu\mu \slashed{E}_T}$. A few tests suggested the cuts give similar results but the more realistic one requires much longer event generation time.
\item $p_{T\ell} > 15$~GeV for both muons and electrons. This is a standard acceptance cut in the LHC experiments. The asymmetries due to new physics increase with an increasing $p_{T\ell}$ cut at the cost of statistical sensitivity. The number we use is a good compromise for million event samples.
\item $|\eta_\ell| < 2.4$ is  a standard acceptance cut for muons and electrons at the LHC experiments and we have checked that this choice does not significantly affect our  asymmetries.
\end{itemize}

In order to preserve the spin correlations it is important to calculate matrix elements for the full process. The numerical implementation of this calculation is significantly complicated by the very narrow $\tau$-lepton width. It is however possible to do it with the current version of Madgraph, which has the necessary numerical precision, at the cost of long event generation times. A simple trick to alleviate this problem is to use a fictitious (and much larger) $\tau$-lepton width during the event generation, and to then rescale the resulting cross-sections by the narrow-width approximation factor $\Gamma_{\tau-fict}/\Gamma_{\tau}$ for each $\tau$ propagator \footnote{We thank Olivier Mattelaer for this suggestion.}, explicitly
\begin{eqnarray}
\sigma&=& \sigma(\Gamma_{\tau-fict}) \times \left(\frac{\Gamma_{\tau-fict}}{\Gamma_{\tau-exp}}\right)^2\times \left(\frac{\Gamma_{\tau-SM}}{\Gamma_\tau(d_{\tau W})}\right)^2
\label{widthfix}
\end{eqnarray}
The first factor is the cross-section calculated by {\tt MadGraph5} using as an input $\Gamma_{\tau-fict}$, typically $2.27\times 10^{-5}$~GeV. The second factor corrects this by  $10^{14}$ by rescaling to the experimental width. Finally, the last factor takes into account the dependence of the $\tau$-width on $d_{\tau W}$. 
To check that this trick does not distort the kinematic distributions of final state leptons to the extent of affecting our asymmetries, we repeated a few calculations using different fictitious values of the $\tau$ width spanning several orders of magnitude. We show the results of this exercise in Table~\ref{tauwidth}.

With the tables presented in the Appendix, we obtain the following approximate numerical fits for the most relevant observables:
\begin{eqnarray}
\frac{\sigma}{\sigma_{SM}}&=&1+ 0.019 \left|d_{\tau W}\right|^2+ 0.0018\left|d_{\tau B}\right|^2+ 0.0053~{\rm Re}(d_{\tau W})  \nonumber \\
A_1 &=& -0.014  ~{\rm Im}(d_{\tau W})  -0.0021 ~{\rm Im}(d_{\tau B})\nonumber \\
A_2 &=& 0.010  ~{\rm Im}(d_{\tau W}) + 0.0023  ~{\rm Im}(d_{\tau B})  \nonumber \\
A_{ss} &=&0.0025  ~{\rm Re}(d_{\tau W})\ {\rm Im}(d_{\tau W})+ 0.031~{\rm Re}(d_{\tau G})\ {\rm Im}(d_{\tau G})+0.031~{\rm Re}(d_{\tau \tilde{G}})\ {\rm Im}(d_{\tau \tilde{G}}) \nonumber \\
A_{C} &=& -0.1125-7.1\times 10^{-4} ~{\rm Re}(d_{\tau W})\nonumber \\
A_{p_T} &=& -0.0955-7.0\times 10^{-4}~{\rm Re}(d_{\tau W})
\label{numfits}
\end{eqnarray}
The salient features of these fits are summarized below.
\begin{itemize}
\item As discussed in Ref.~\cite{Hayreter:2013vna}, terms in the cross-section linear in the real part of the anomalous couplings are suppressed by the $\tau$ mass at LHC energies and this is confirmed both by our fit and by the symmetry of Figure~\ref{bsigma}.
 Our fits are not quite the same as the ones we presented in  Ref.~\cite{Hayreter:2013vna} due to the different $p_{T\ell}$ cut used there and the different normalization for $d_{\tau W}$. With the cuts used here, the bounds placed on the anomalous couplings by measurements of the cross-section  (using the same procedure as in Ref.~\cite{Hayreter:2013vna}) are shown in Figure~\ref{bsigma}. 
\begin{figure}[h]
\includegraphics[scale=0.65]{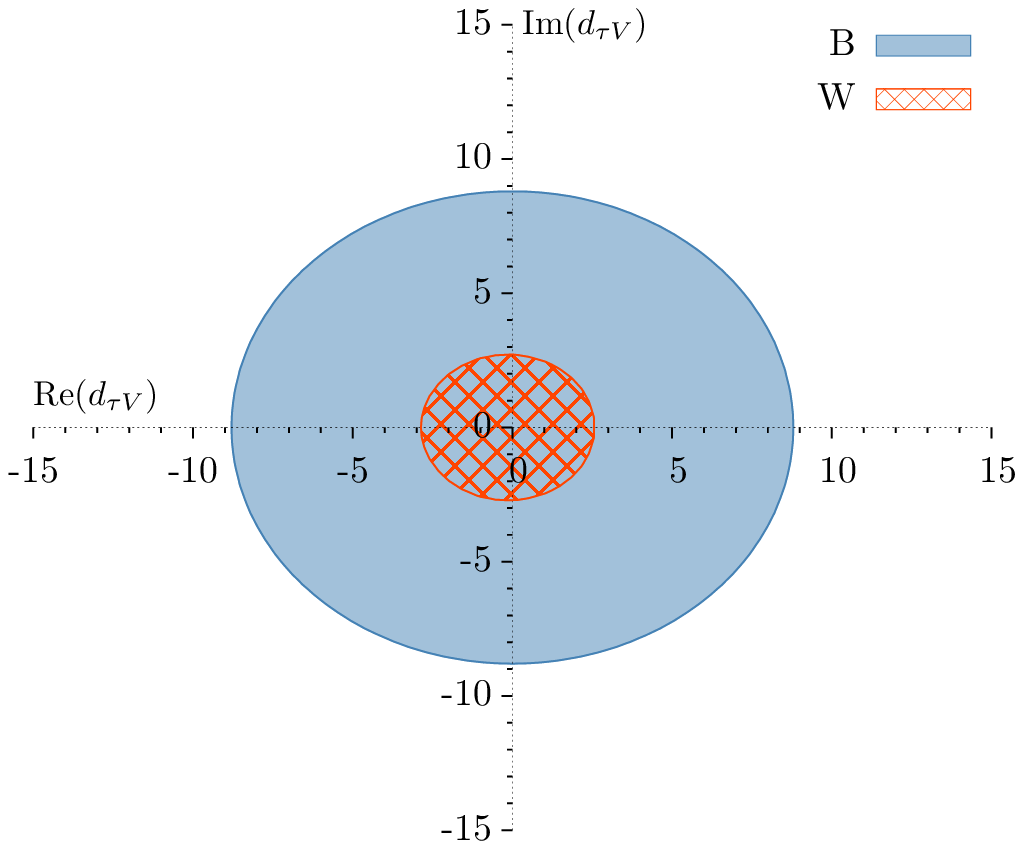} \hspace*{0.2cm}
\includegraphics[scale=0.65]{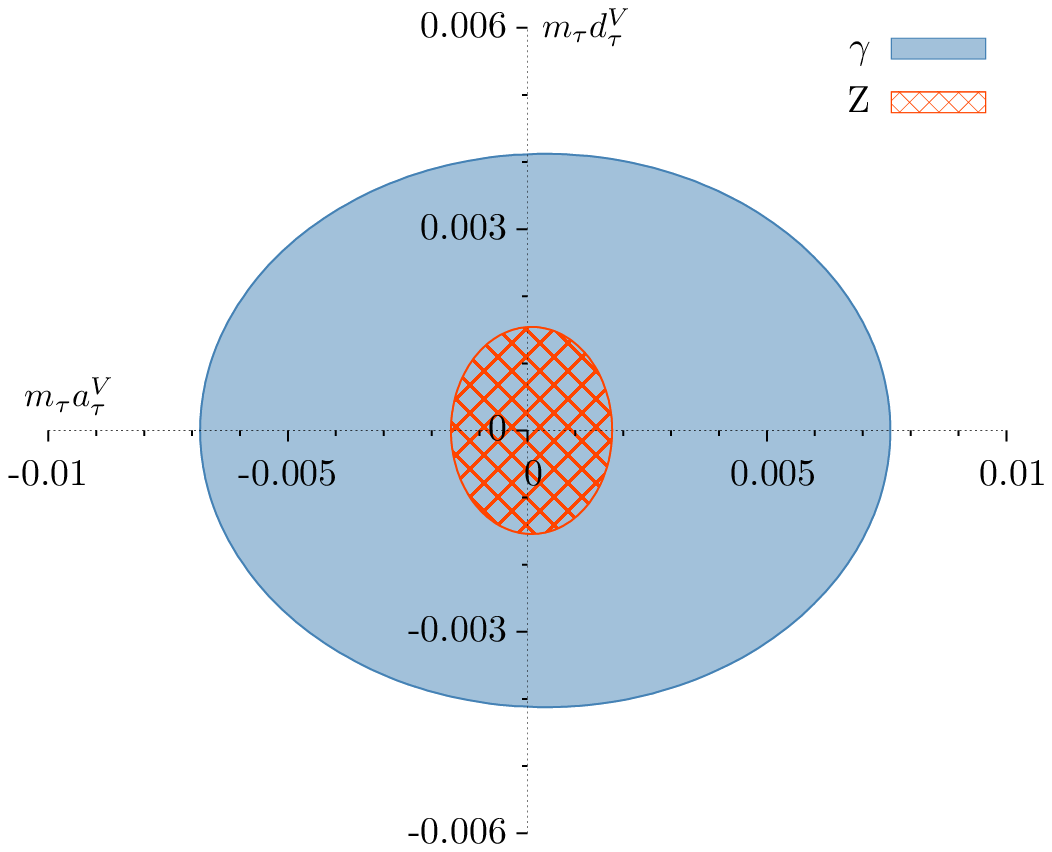} 
\caption{Regions of $d_{\tau V}$ (left) and the corresponding $d_{\tau}^{ \gamma,Z}$, $a_{\tau}^{ \gamma,Z}$ (right) allowed by a maximum 14\% deviation from the SM cross-section with the cuts described in the text.}
\label{bsigma}
\end{figure}   
Taking only one parameter to be non-zero at a time we find,
\begin{eqnarray}
|{\rm Im}(d_{\tau W})| \lsim 2.7 &,& |{\rm Im}(d_{\tau B})| \lsim 8.8 \nonumber \\
-2.9  \lsim {\rm Re}(d_{\tau W}) \lsim 2.6 &,&
 |{\rm Re}(d_{\tau B})|\lsim 8.8
\end{eqnarray}
or equivalently,
\begin{eqnarray}
|m_\tau d_{\tau}^Z| \lsim 0.0015 &,& |m_\tau d_{\tau}^\gamma| \lsim 0.004 \nonumber \\
-0.0016  \lsim m_\tau a_{\tau}^Z \lsim 0.0018 &,& -0.0068  \lsim  m_\tau a_{\tau}^\gamma \lsim 0.0076
\end{eqnarray}
These constraints are similar to those quoted in Table~\ref{tau:results}, which is not surprising because in both cases they  correspond to assuming that the cross-section will be measured to 14\% accuracy.\footnote{This precision corresponds to the largest systematic error in the CMS analysis of high invariant mass $\tau$-pairs \cite{Chatrchyan:2012hd}}.

\item A glance at the Feynman diagrams for $q\bar{q} \to \tau^+\tau^- \to \mu^+\mu^- \nu_\tau \bar\nu_\tau \nu_\mu \bar\nu_\mu$ reveals that the cross-section should be quadratic in $d_{\tau B}$ and a polynomial of order 6 in $d_{\tau W}$ because the latter also appears in the $\tau$ decay vertex as implied by the gauge invariant form of the operators, Eq.~\ref{ginvedm}.  Our numerical calculation indicates that the cross-section has a sensitivity to $d_{\tau W}$ at most quadratic, in other words the precision of our simulations makes it difficult to allow for the higher order terms. This is because our procedure is a form of the narrow width approximation (but keeping spin correlations): the dependence of $\sigma(pp\to \tau^+\tau^-)$ on $d_{\tau W}$ is quadratic, and the $\tau$-lepton branching ratios remain approximately independent of $d_{\tau W}$. 

Interestingly, the $\tau$-width itself depends on $d_{\tau W}$ and we could use that to find an additional constraint. In the approximation in which we treat the hadronic $\tau$-decay as decay into free quarks, we find
\begin{eqnarray}
\Gamma_\tau(d_{\tau W}) \approx \Gamma_{\tau-SM}\left(1+0.00126{\rm ~Re}(d_{\tau W}) +\cdots\right)
\label{fixe}
\end{eqnarray}
where $\cdots$ stands for much smaller quadratic corrections, and this is the precise factor we use in Eq.~\ref{widthfix}. Of course this approximation does not calculate  the hadronic decay modes correctly, but it gives us an estimate\footnote{Recall that this approximation used for the SM results in a $\tau$-width that is only about 10\% smaller than the experimental width.} for the size of the corrections introduced by $d_{\tau W}$.  Taken literally, and using  from the particle data book that the $\tau$ mean life is $(290.3\pm 0.5)\times 10^{-15}$s \cite{Agashe:2014kda}, it implies
\begin{equation}
|{\rm Re}(d_{\tau W})|\lsim 1.4
\end{equation}

\item The $T$-odd correlations $A_{1,2}$ exhibit a linear dependence on the imaginary part of the anomalous couplings that is not suppressed by the $\tau$ mass, as is expected for single spin correlations. We find that this process is about six times more sensitive to ${\rm Im}(d_{\tau W})$ than to ${\rm Im}(d_{\tau B})$. 
\item The $T$-odd and $CP$-odd asymmetry $A_{ss}$ receives contributions quadratic in the anomalous couplings that are not suppressed by the $\tau$ mass. As discussed above, they arise from double spin asymmetries produced in the interference of the new physics amplitudes with themselves.
\item The independence on the $\tau$-mass for two  asymmetries originating in the interference between new physics ${\rm Im}(d_{\tau W}) =10$ and the SM (since ${\rm Re}(d_{\tau W}) =0$) is shown in Table~\ref{taumass}.
\item The $T$ and $CP$-even asymmetries $A_{C,p_T}$ exhibit the linear dependence on the real part of the anomalous couplings implied by the single spin correlation. 
\item At the level of our study, the final dimuon channel can be replaced with the  dilepton channel (including muons and electrons) and this increases the statistics by a factor of four without affecting the asymmetries. We explicitly compute the asymmetries for the dilepton channel for one value of $d_{\tau W}$ in Table~\ref{muande} obtaining the same asymmetries as in the dimuon channel given in the other Tables. The asymmetries in the dilepton channel are generalized from the dimuon case based on the lepton charge  (so we also include the $\mu^+e^-$ and $\mu^-e^+$ final states).
\end{itemize}

The statistical sensitivity to any of the asymmetries in the dilepton channel with 100 fb$^{-1}$ is 0.005. This translates into the following future constraints
\begin{eqnarray}
|{\rm Im}(d_{\tau W})| \lsim 0.36 &,& |{\rm Im}(d_{\tau B})| \lsim 2.2 \nonumber \\
|m_\tau d_{\tau}^Z| \lsim 2\times 10^{-4} &,& |m_\tau d_{\tau}^\gamma| \lsim 1\times 10^{-3} \nonumber \\
|{\rm Re}(d_{\tau W})| \lsim 7 &,& 
| m_\tau a_\tau^Z| \lsim 0.0043, \,\,\,  |m_\tau a_\tau^\gamma| \lsim 0.019 \nonumber \\
|{\rm Re}(d_{\tau W}){\rm Im}(d_{\tau W})| \lsim 2 &,& 
|{\rm Re}(d_{\tau G, \tilde{G}}){\rm Im}(d_{\tau G, \tilde{G}})| \lsim 0.16
\label{1sigmabounds}
\end{eqnarray}
It is worth noting here that whether we include Eq.\ref{fixe} or not in Eq.\ref{widthfix} only affects the constraint we find for ${\rm Re}(d_{\tau W})$, increasing it to 7 from 5.

\subsection{Dilepton pairs in the $Z$-resonance region}

The $Z$-resonance region is selected with the cut $60 < m_{\tau\tau} < 120$~GeV, with the same caveats as before. Keeping the remaining cuts unchanged and generating additional samples we obtain the following approximate fits.
\begin{eqnarray}
\frac{\sigma_Z}{\sigma_{SM}}&=&1+ 0.0069\ {\rm Re}(d_{\tau W})^2+ 0.0066\ {\rm Im}(d_{\tau W})^2+0.0006\left|d_{\tau B}\right|^2+ 0.0024~{\rm Re}(d_{\tau W})  \nonumber \\
A_1 &=& -0.013 ~{\rm Im}(d_{\tau W}) -0.0038 ~{\rm Im}(d_{\tau B}) \nonumber \\
A_2 &=& 0.0016 ~{\rm Im}(d_{\tau W}) +0.0005 ~{\rm Im}(d_{\tau B})\nonumber \\
A_{ss} &=& 0.0011 ~{\rm Re}(d_{\tau W})\ {\rm Im}(d_{\tau W})+0.001~{\rm Re}(d_{\tau G})\ {\rm Im}(d_{\tau G})+0.001~{\rm Re}(d_{\tau \tilde{G}})\ {\rm Im}(d_{\tau \tilde{G}})\nonumber \\
A_{C} &=& -0.0279-5.4\times 10^{-4}~{\rm Re}(d_{\tau W})\nonumber \\
A_{p_T} &=& -0.0251-4.7\times 10^{-4}~{\rm Re}(d_{\tau W}) 
\end{eqnarray}
We have not included ${\rm Re}(d_{\tau B})$ for the asymmetries due to the smaller sensitivity already observed in the previous case. Our main observations in this case are:

\begin{itemize}

\item Constraints arising from the cross-section are shown in Figure~\ref{bsigz}. In this case we have assumed the cross-section can be measured to 7\% accuracy, the current systematic uncertainty, following  Ref.~\cite{Aad:2014jra}. 
\begin{figure}[h]
\includegraphics[scale=0.65]{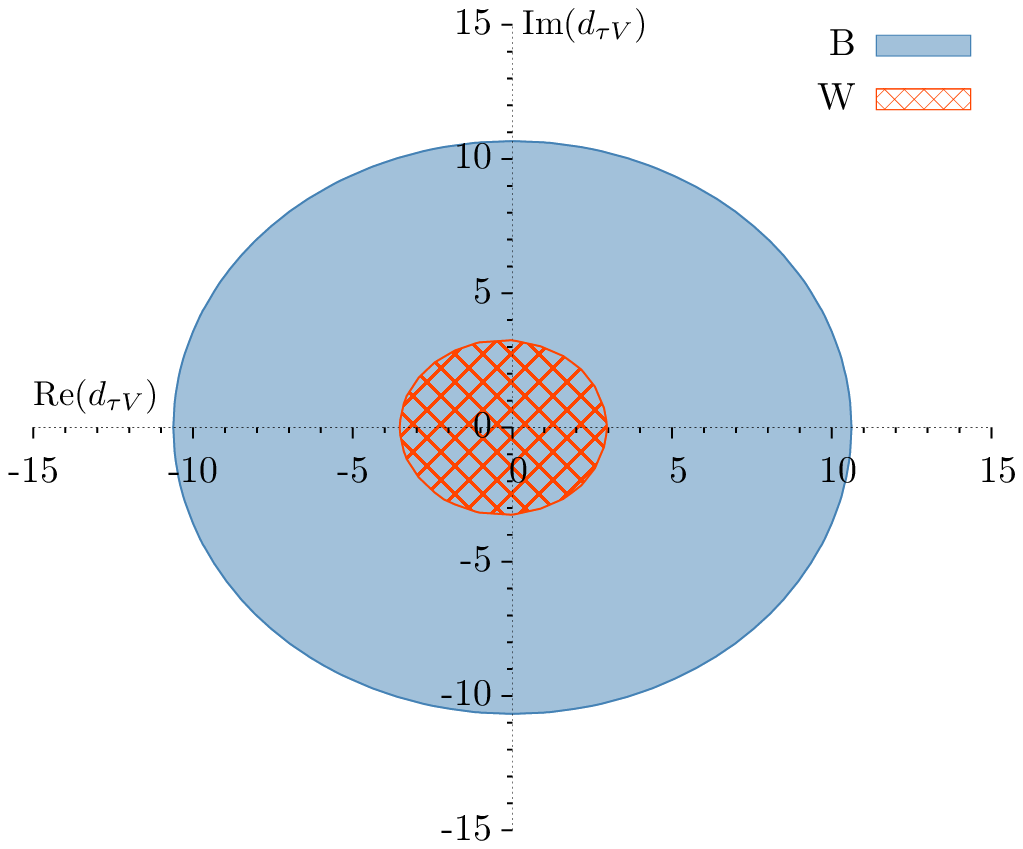} \hspace*{0.2cm}
\includegraphics[scale=0.65]{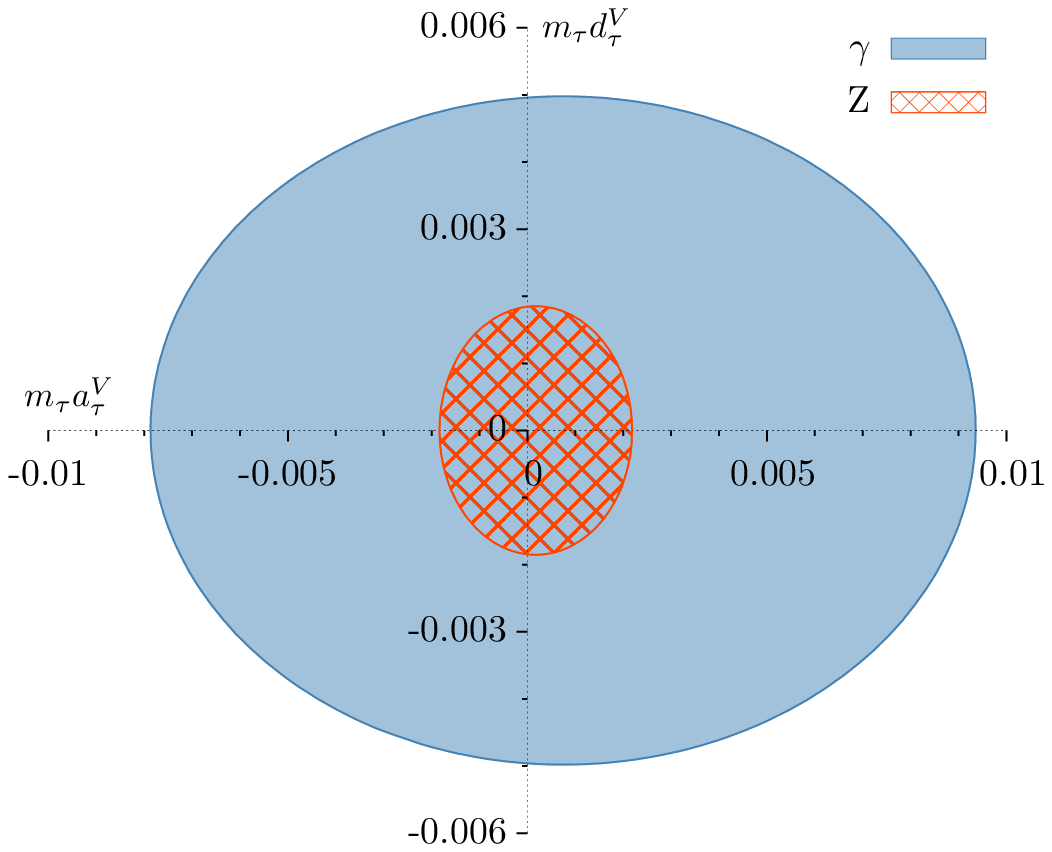} \\
\caption{Regions of $d_{\tau V}$ (left) and the corresponding $d_{\tau}^{ \gamma,Z}$, $a_{\tau}^{ \gamma,Z}$ (right) allowed by a maximum 7\% deviation from the SM cross-section with the cuts described in the text. }
\label{bsigz}
\end{figure}   
Taking only one parameter to be non-zero at a time we find,
\begin{eqnarray}
|{\rm Im}(d_{\tau W})| \lsim 3.3 &,& |{\rm Im}(d_{\tau B})| \lsim 10.7 \nonumber \\
-3.5  \lsim {\rm Re}(d_{\tau W}) \lsim 3.0 &,&
|{\rm Re}(d_{\tau B})|\lsim 10.6
\end{eqnarray}
or equivalently,
\begin{eqnarray}
|m_\tau d_{\tau}^Z| \lsim 0.0018 &,& |m_\tau d_{\tau}^\gamma| \lsim 0.005 \nonumber \\
-0.0018  \lsim m_\tau a_{\tau}^Z \lsim 0.0021 &,& -0.0078  \lsim  m_\tau a_{\tau}^\gamma \lsim 0.0093
\end{eqnarray}
which are between 20-25\% weaker than those that can be obtained from high energy pairs and at best comparable to the existing constraints from LEP.

\item Assuming again 100 fb$^{-1}$ implies a much better statistical sensitivity of 0.0009 when using both electron and muon channels. This better sensitivity is of course due to the much larger cross-section and results in the potential constraints
\begin{eqnarray}
|{\rm Im}(d_{\tau W})| \lsim 0.07 &,& |{\rm Im}(d_{\tau B})| \lsim 0.24 \nonumber \\
|m_\tau d_{\tau}^Z| \lsim 4.3 \times 10^{-5} &,& |m_\tau d_{\tau}^\gamma| \lsim 1.9\times 10^{-4} \nonumber \\
|{\rm Re}(d_{\tau W})| \lsim 1.7 &,& |m_\tau a_{\tau}^Z| \lsim 0.0011 ~,~ |m_\tau a_{\tau}^\gamma| \lsim 0.0045 \nonumber \\
|{\rm Re}(d_{\tau W}) {\rm Im}(d_{\tau W})| \lsim 0.82 &,& 
|{\rm Re}(d_{\tau G, \tilde{G}}){\rm Im}(d_{\tau G, \tilde{G}})| \lsim 0.9
\end{eqnarray}
\end{itemize}

\subsection{Background}

We end this section with a brief discussion of background and how it would affect the constraints estimated so far.  For the dilepton channel in $\tau$-pair production both $\tau$-leptons in the pair undergo leptonic decay into muons or electrons:  $pp\to \tau^+\tau^- \to \ell^+ \ell^- \slashed{E}_T$, $\ell =\mu,e$ where the missing transverse energy, $\slashed{E}_T$ is due to invisible neutrinos. If one lepton is a muon and the other one an electron the dominant background arises from $t\bar{t}$, $W^+ W^-$ or $ZZ$ production. If the two leptons have the same flavor there is an additional direct Drell-Yan production of $\ell^+\ell^-$. In addition, as discussed for example in Ref.~\cite{Aad:2012gm}, contributions from processes where a jet or a photon is misidentified as a lepton are very small.

The different handles to control this background have been identified by the experimental collaborations. Requiring a minimum missing $E_T$ can effectively remove the direct Dell-Yan background, we will use $\slashed{E}_T > 20$~GeV. To suppress $t\bar{t}$ background experiments require at most one jet and no $b$ tagged jets. These requirements are hard to implement at the level of our analysis but they should be kept in mind. The requirement that the two leptons be back to back in the transverse plane provides additional suppression against top-pairs and $W$ and $Z$ pairs. This is implemented as \cite{Chatrchyan:2012hd}
\begin{eqnarray}
\cos\Delta\phi(\ell^-,\ell^+)<-0.95
\end{eqnarray}
where $\Delta\phi(\ell^-,\ell^+)$ is the difference in azimuthal angle between lepton pairs. And to further suppress the contamination from $W$ products, events are selected with an additional requirement that the signature being consistent with that of a particle decaying into two $\tau$ leptons. With the following projection variables \cite{Chatrchyan:2012hd}
\begin{eqnarray}
p^{vis}_{\xi}&=&\vec{p}_{T\ell^+}\cdot \hat{\xi}+\vec{p}_{T\ell^-}\cdot \hat{\xi}, \nonumber \\
p_\xi&=&p^{vis}_{\xi}+\overrightarrow{E_T^{\rm miss}}\cdot \hat{\xi}
\end{eqnarray}
we require $p_\xi-(1.25\times p^{vis}_\xi) > -10$, where $\hat{\xi}$ is a unit vector along the bisector of the momenta of the two leptons.

Generating MonteCarlo samples for each of the background processes with Madgraph and applying all the preceding cuts to background and signal samples results in cross sections:

\begin{eqnarray}
\sigma(pp \to \tau^+\tau^- \to \mu^+\mu^- \slashed{E}_T) &=& 20.28~{\rm fb} \nonumber \\
\sigma(pp \to t \bar{t} \to b \bar{b} \mu^+\mu^- \slashed{E}_T) &=& 120.1~{\rm fb} \nonumber \\
\sigma(pp \to W^+ W^- \to \mu^+\mu^- \slashed{E}_T)  &=& 18.16~{\rm fb} \nonumber \\
\sigma(pp \to Z Z \to \mu^+\mu^- \slashed{E}_T)  &=& 1.75~{\rm fb}
\end{eqnarray}
These numbers allow us to quantify the effect of background as follows. First, the cuts needed to isolate the signal reduce its cross-section by a factor of 4.7 which results in a factor of 2.2 loss in statistical sensitivity. In addition the final event sample will contain background events that, under our assumptions, are not affected by the new physics. If we use the CMS \cite{Chatrchyan:2012jua} b-tagging efficiency between 70-85\% we expect less than 11~fb of $\sigma(pp \to t \bar{t} )$ background to remain. In this case the T-odd 
asymmetries are reduced by about 2.5. 

Bounds estimated from the cross-sections do not depend so much on the background cross-section as on its uncertainty and this has already been taken into account when we use the experimental estimates for the precision they can achieve in their cross-section measurements. The effect of background on the T-even asymmetries is much harder to estimate, but these do not improve the bounds on ${\rm Re}(d_{\tau V})$ significantly over bounds obtained from cross-sections in any case.

\section{Summary}

We have examined the possible limits that can be placed on certain anomalous couplings of $\tau$-leptons at the LHC14 with 100~fb$^{-1}$. We have considered the four dipole-type couplings that appear at dimension six in the effective Lagrangian as well as the two $\tau$-gluon couplings that appear at dimension eight. We have found the statistical sensitivity of single and double spin asymmetries in the dilepton channel to these couplings and compared them to the statistical sensitivity from measuring deviations from the SM cross-section at the 14\% level. We find that $T$ odd asymmetries can improve the bounds on the $CP$ violating couplings but that single-spin asymmetries do not seem to improve the bounds on the anomalous magnetic moments.

\begin{acknowledgments}

This work was supported in part by the DOE under contract number DE-SC0009974. We thank David Atwood for useful discussions. G.V. thanks the theory group at CERN for their hospitality and partial support while this work was completed.

\end{acknowledgments}

\appendix

\section{Tables}

All the tables are produced from million event samples for the process $pp\to \tau^+\tau^-\to \mu^+\mu^-\nu_\tau \bar{\nu}_\tau\nu_\mu\bar{\nu}_\mu$ at 14~TeV obtained with the following cuts: $p_{T\mu}\geq 15$~GeV, $|\eta|_\mu \leq 2.4$, $m_{\tau\tau}>120$~GeV. With the exceptions noted explicitly below, the $\tau$ width was set to $2.27\times 10^{-5}$~GeV and the resulting cross-sections were then scaled as described in the main text.

\begin{table}[h]
\begin{center}
\begin{tabular}{|c|c|c|c|c|c|c|c|}\hline
Re($d_{\tau W}$) & Im($d_{\tau W}$) & $\sigma (\rm fb)$ & $A_1$ & $A_2$ & $A_{test}$ & $A_{C}$ & $A_{p_T}$ \\\hline
0 & 0 & 95.45 & 0.0003 & 0.0000 & 0.0019 & -0.1125 & -0.0955\\\hline
0 & 2 & 102.7 & -0.0283 & 0.0204 & -0.0002 & -0.1133 & -0.0956\\\hline
0 & 4 & 124.3 & -0.0579 & 0.0399 & -0.0013 & -0.1119 & -0.0962\\\hline
0 & 6 & 160.4 & -0.0858 & 0.0613 & -0.0018 & -0.1134 & -0.0938\\\hline
0 & 8 & 210.9 & -0.1117 & 0.0780 & 0.0022 & -0.1088 & -0.0965\\\hline
0 & 10 & 276.0 & -0.1457 & 0.1019 & 0.0013 & -0.1131 & -0.0991\\\hline
2 & 0 & 103.7 & -0.0003 & -0.0013 & 0.0009 & -0.1141 & -0.0964\\\hline
2 & 2 & 111.0 & -0.0294 & 0.0194 & 0.0013 & -0.1170 & -0.0985\\\hline
2 & 4 & 132.6 & -0.0579 & 0.0384 & 0.0007 & -0.1175 & -0.0988\\\hline
2 & 6 & 168.6 & -0.0871 & 0.0622 & -0.0006 & -0.1150 & -0.0964\\\hline
2 & 8 & 219.4 & -0.1145 & 0.0797 & -0.0008 & -0.1171 & -0.1026\\\hline
2 & 10 & 284.2 & -0.1453 & 0.1003 & -0.0012 & -0.1206 & -0.1029\\\hline
4 & 0 & 126.4 & -0.0026 & 0.0011 & -0.0007 & -0.1155 & -0.1000\\\hline
4 & 2 & 133.6 & -0.0284 & 0.0212 & -0.0003 & -0.1178 & -0.1019\\\hline
4 & 4 & 155.3 & -0.0565 & 0.0380 & 0.0003 & -0.1177 & -0.1009\\\hline
4 & 6 & 191.4 & -0.0837 & 0.0640 & -0.0014 & -0.1152 & -0.0986\\\hline
4 & 8 & 242.0 & -0.1140 & 0.0820 & -0.0022 & -0.1175 & -0.1012\\\hline
4 & 10 & 306.9 & -0.1445 & 0.0971 & 0.0022 & -0.1186 & -0.1027\\\hline
\end{tabular}
\end{center}
\caption{Single spin $T$-odd correlations $A_{1,2}$ and $T$-even correlations $A_{C,p_T}$ for several values of ${\rm Re}(d_{\tau W})$ and ${\rm Im}(d_{\tau W})$. $A_{test}$ should vanish in all cases and gives us an estimate of the statistical error. }
\label{trwiw}
\end{table}

\begin{figure}[h]
\includegraphics[width=0.45\textwidth]{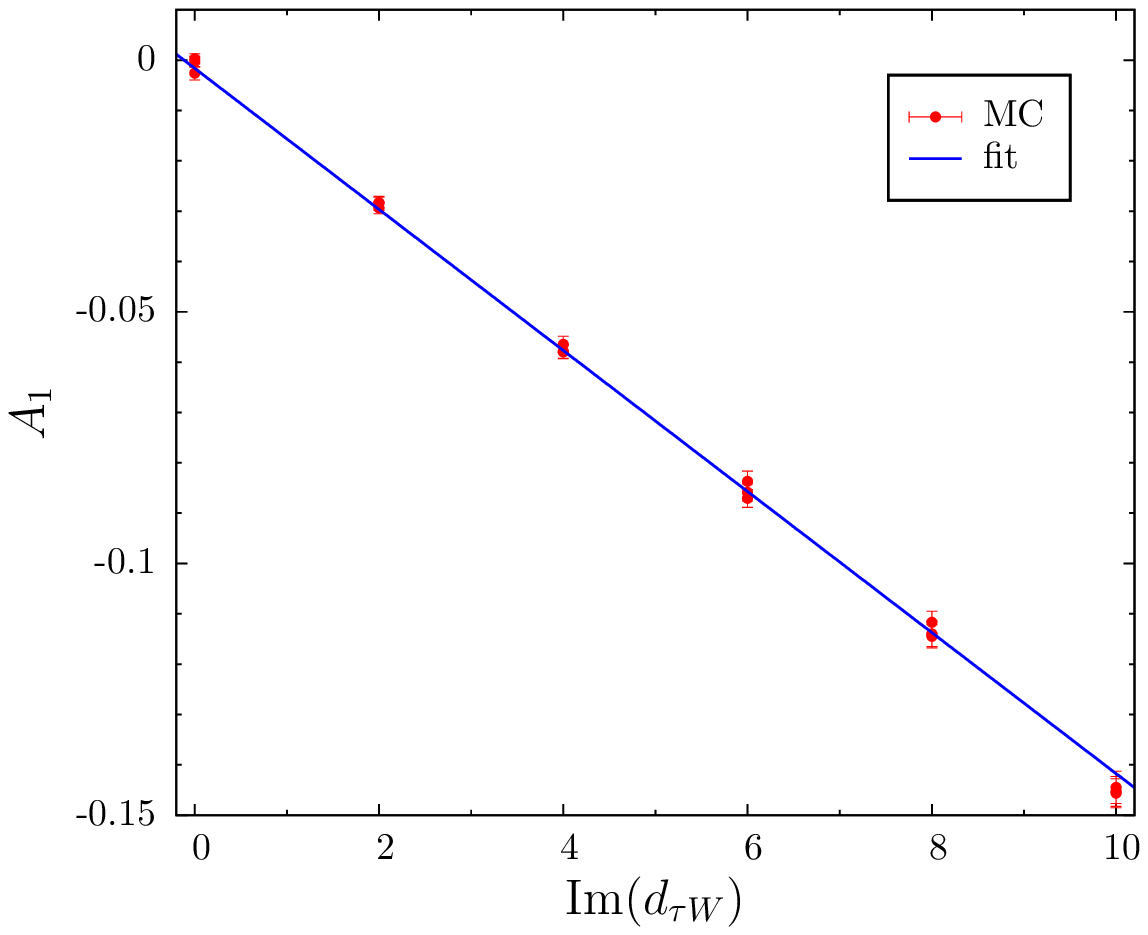} \hspace{0.5in}
\includegraphics[width=0.45\textwidth]{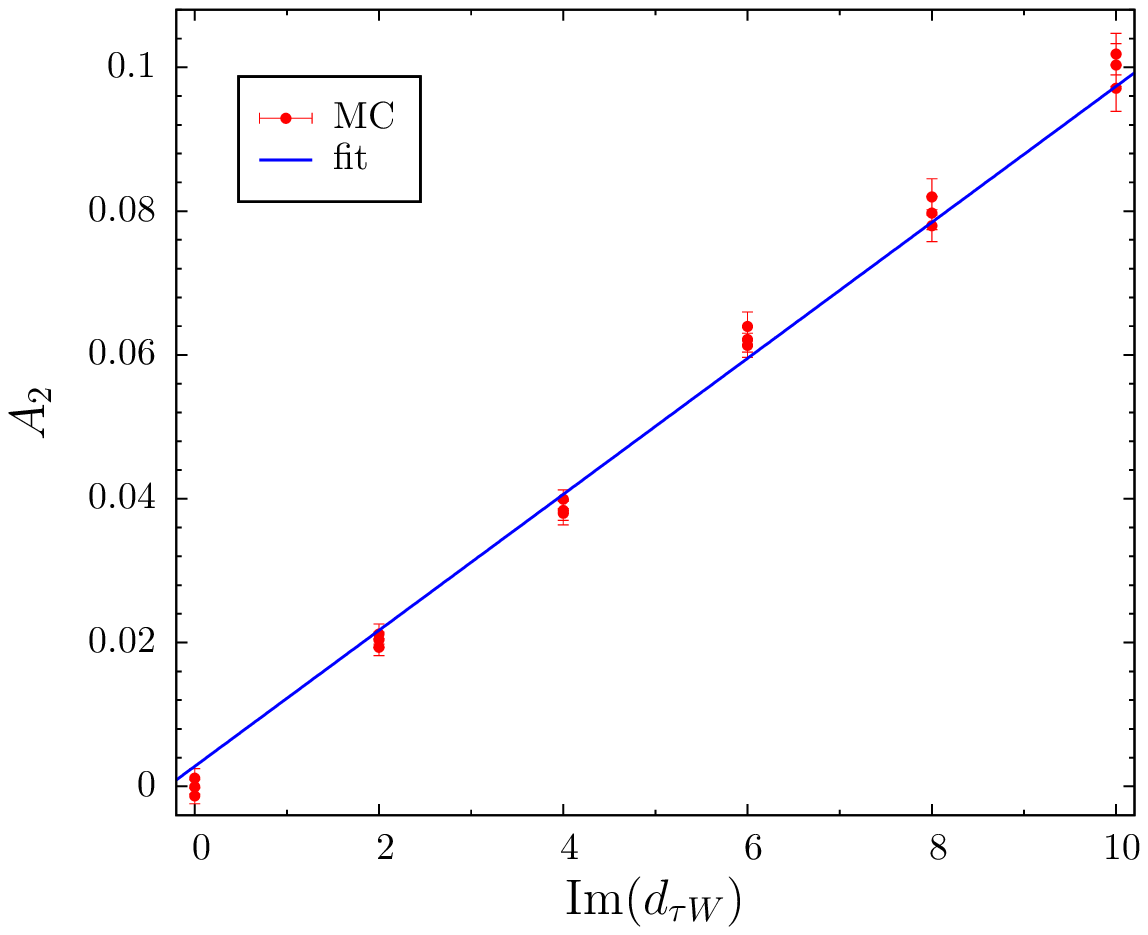}
\caption{A linear fit to  ${\rm Im}(d_{\tau W})$ for $A_1$ (left) and  $A_2$ (right) in Table~\ref{trwiw} is supported by the data. The separation between the three points corresponding to three values of ${\rm Re}(d_{\tau W})$ suggests a very small contribution of the form ${\rm Re}(d_{\tau W}){\rm Im}(d_{\tau W})$.}
\label{fwir45}
\end{figure}

\begin{figure}[h]
\includegraphics[width=0.45\textwidth]{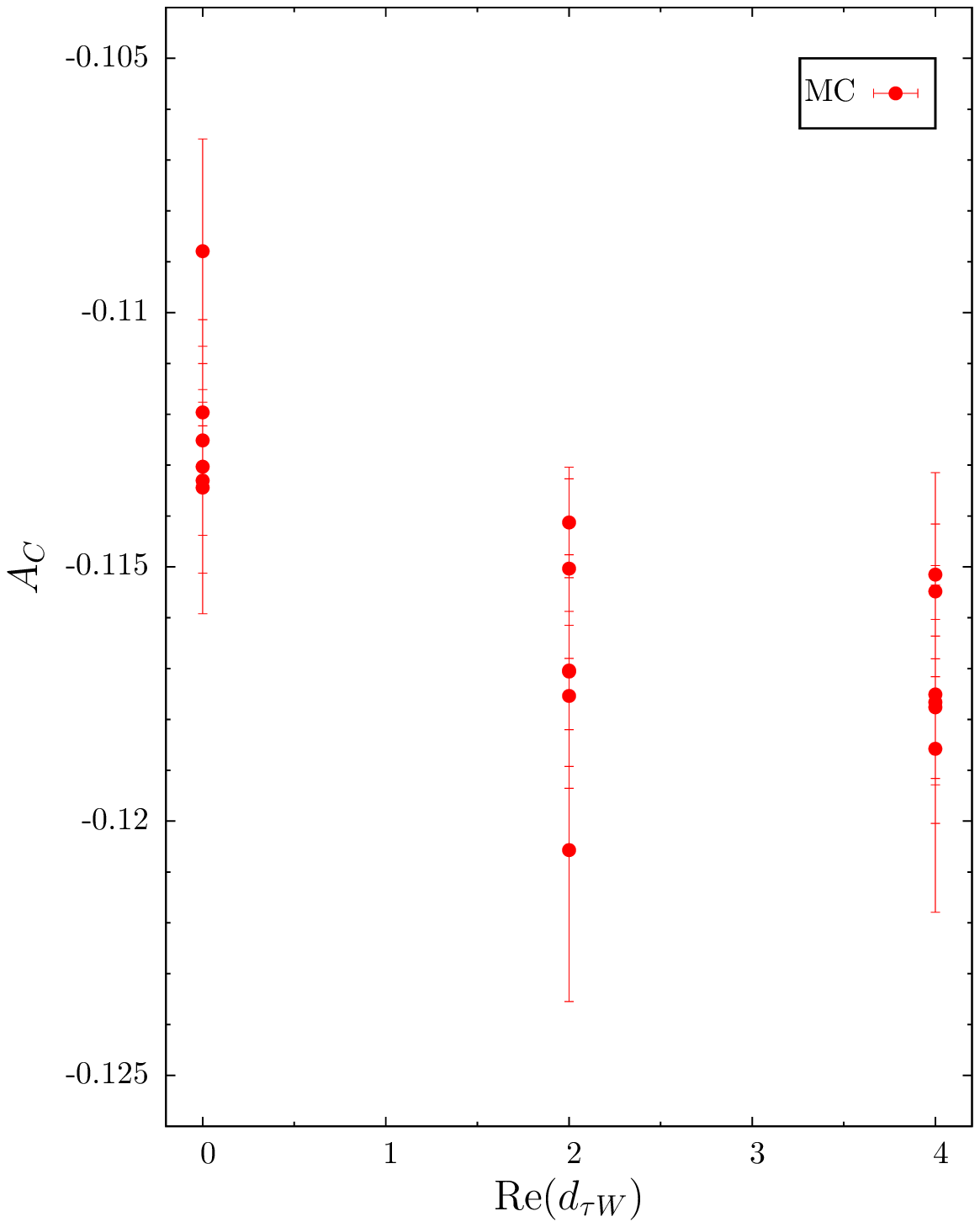} \hspace{0.5in}
\includegraphics[width=0.45\textwidth]{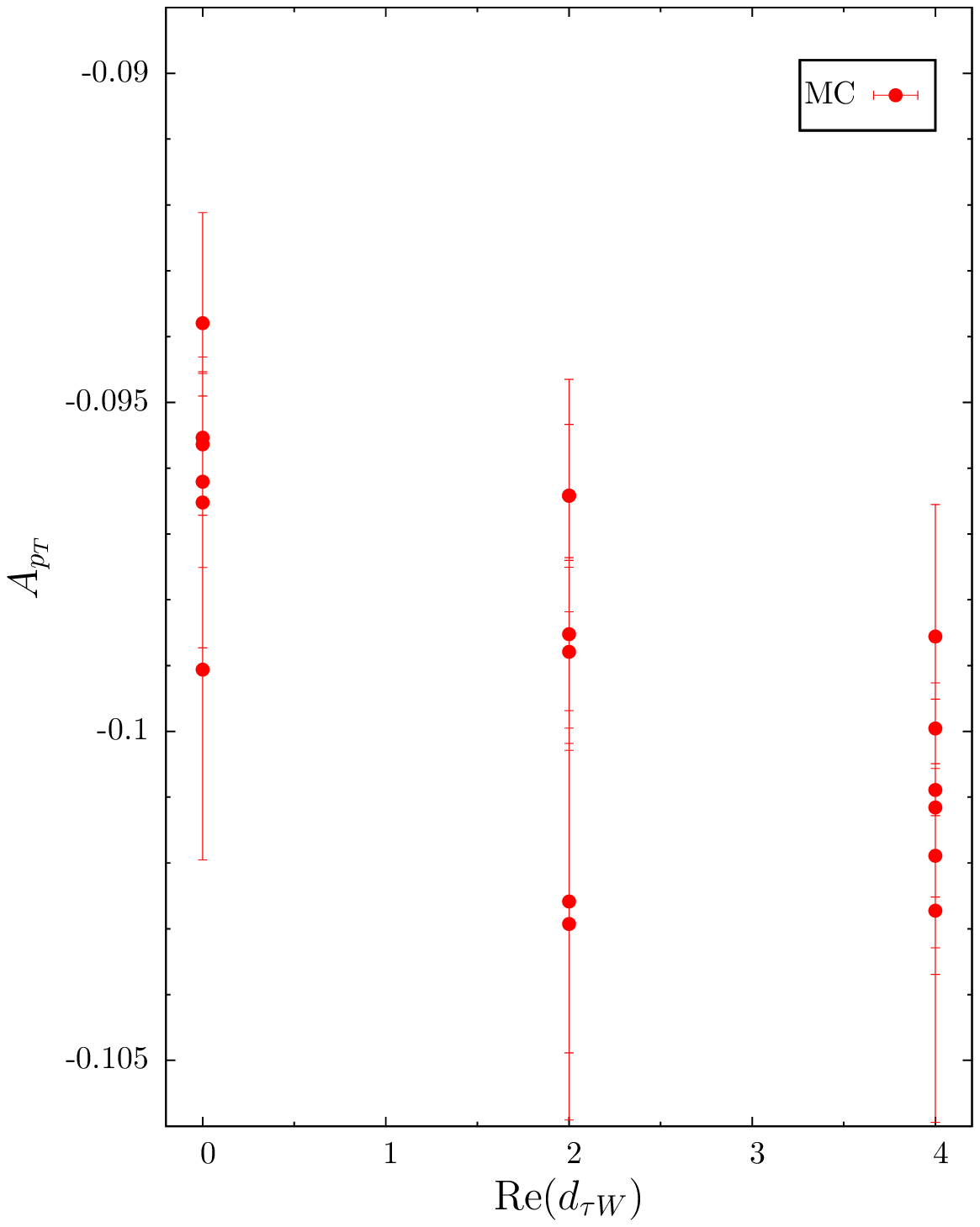}
\caption{Values of $A_{C}$ (left) and  $A_{p_T}$ (right) in Table~\ref{trwiw}. The separation between the five points corresponding to the values of ${\rm Im}(d_{\tau W})$ for each ${\rm Re}(d_{\tau W})$ suggests sizeable contributions of the form $({\rm Im}(d_{\tau W}))^2$.}
\label{fwir1011}
\end{figure}

In Table~\ref{trwiw} we compute the single-spin asymmetries chosen above for a series of values of ${\rm Re}(d_{\tau W})$ and ${\rm Im}(d_{\tau W})$ along with $A_{test}$ which should be zero up to statistical error. In Figure~\ref{fwir45} we plot the $T$-odd asymmetries which exhibit the expected behavior linear in ${\rm Im}(d_{\tau W})$. The figure also suggests that they have very small dependence on ${\rm Re}(d_{\tau W})$. In Figure~\ref{fwir1011} we plot the $T$-even asymmetries, the situation is less clear in this case and we need to study other tables to reach any conclusions.

\begin{figure}[h]
\includegraphics[width=0.45\textwidth]{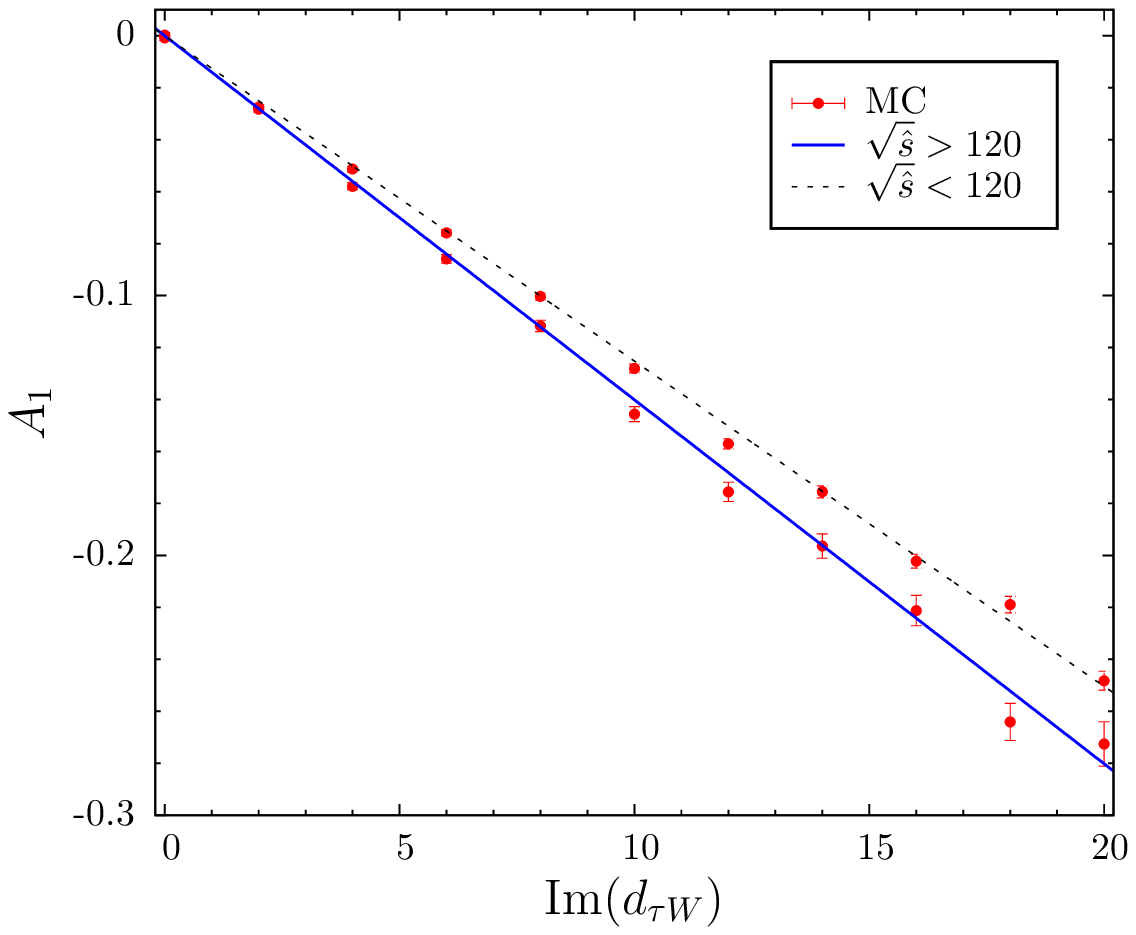} \hspace{0.5in}
\includegraphics[width=0.45\textwidth]{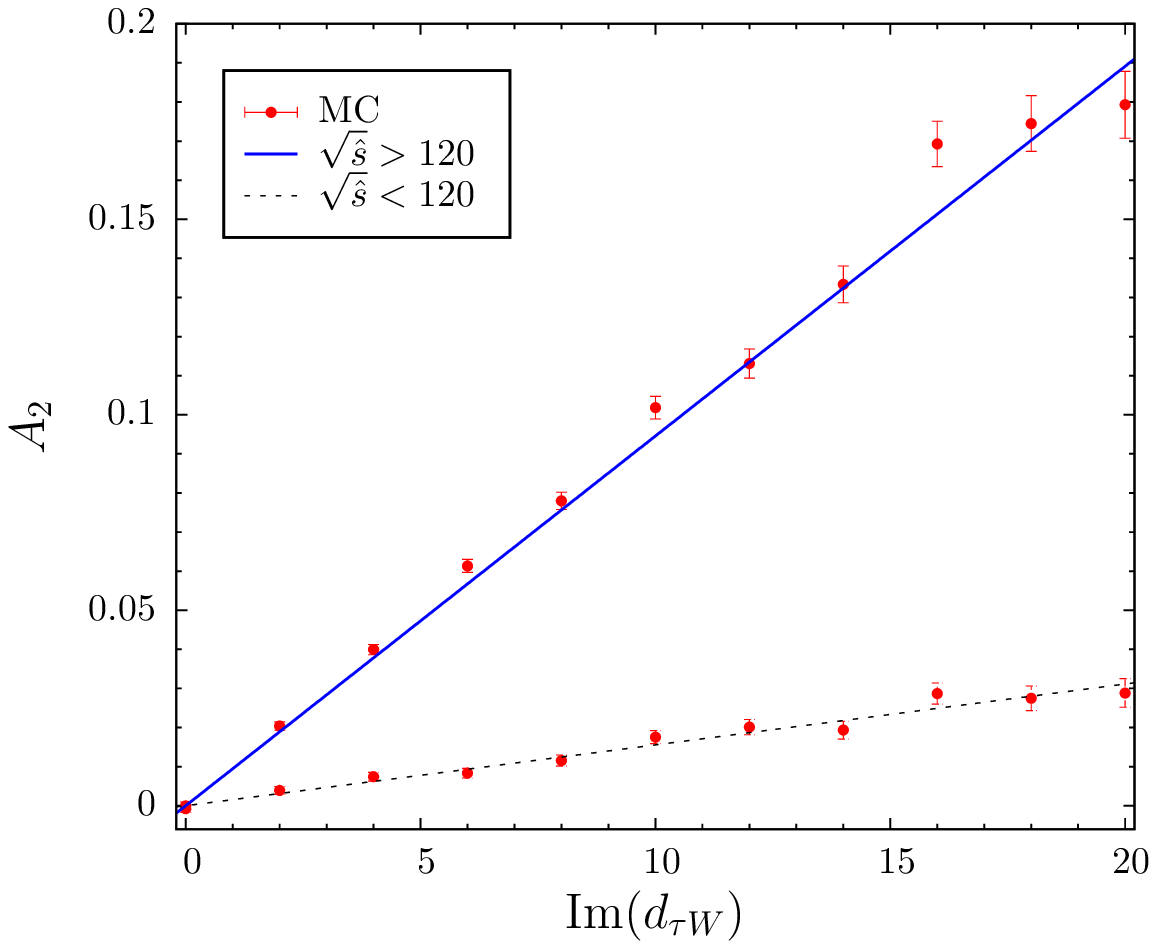} 
\caption{MC simulation compared to a linear fit to  ${\rm Im}(d_{\tau W})$ for $A_1$ (left) and  $A_2$ (right) for high $m_{\tau\tau}$ events from Table~\ref{tiw} and also for events in the $Z$-region $60<m_{\tau\tau}<120$~GeV .}
\label{fiw45}
\end{figure}

\begin{figure}[h]
\includegraphics[width=0.45\textwidth]{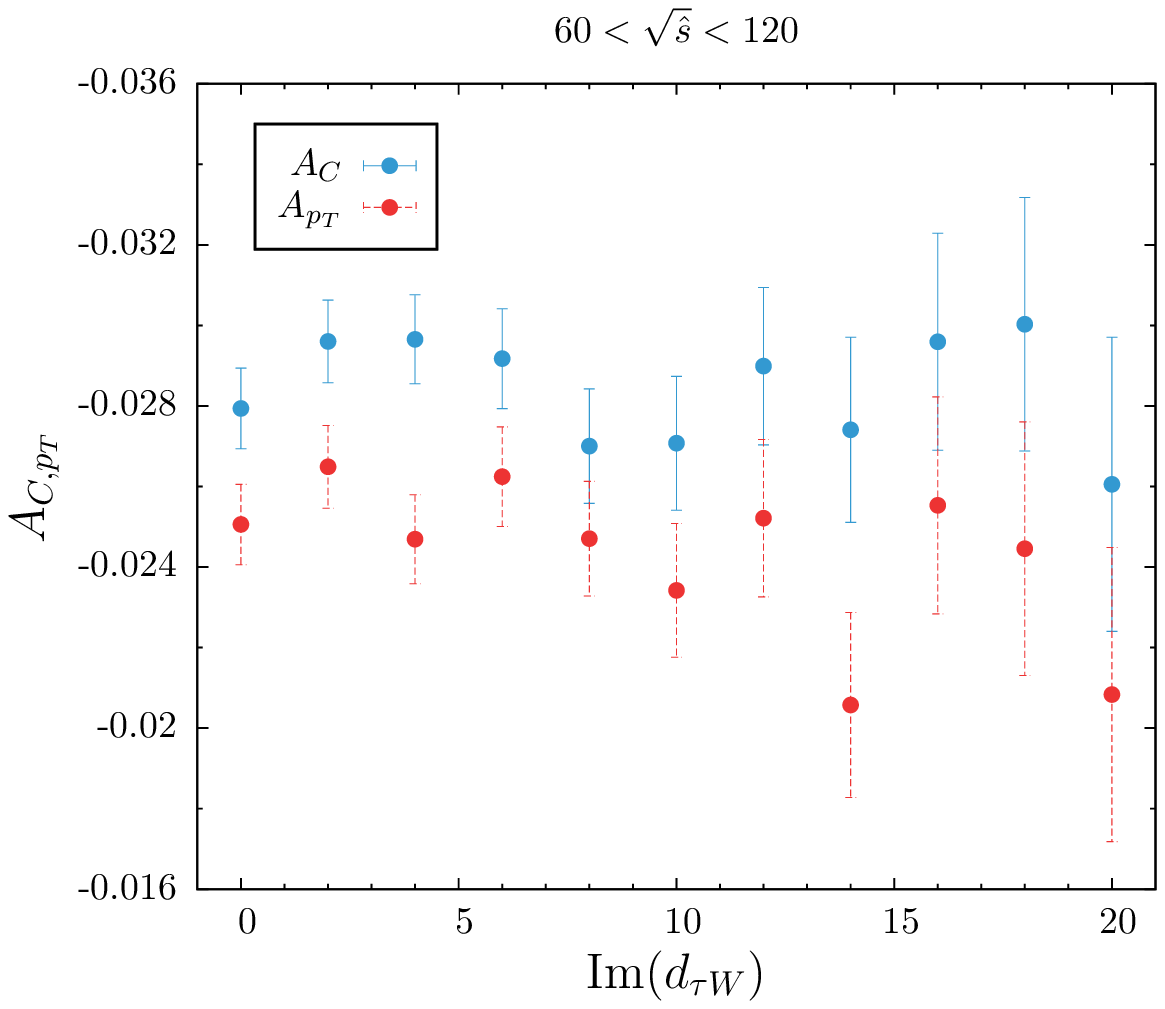} \hspace*{0.5cm}
\includegraphics[width=0.45\textwidth]{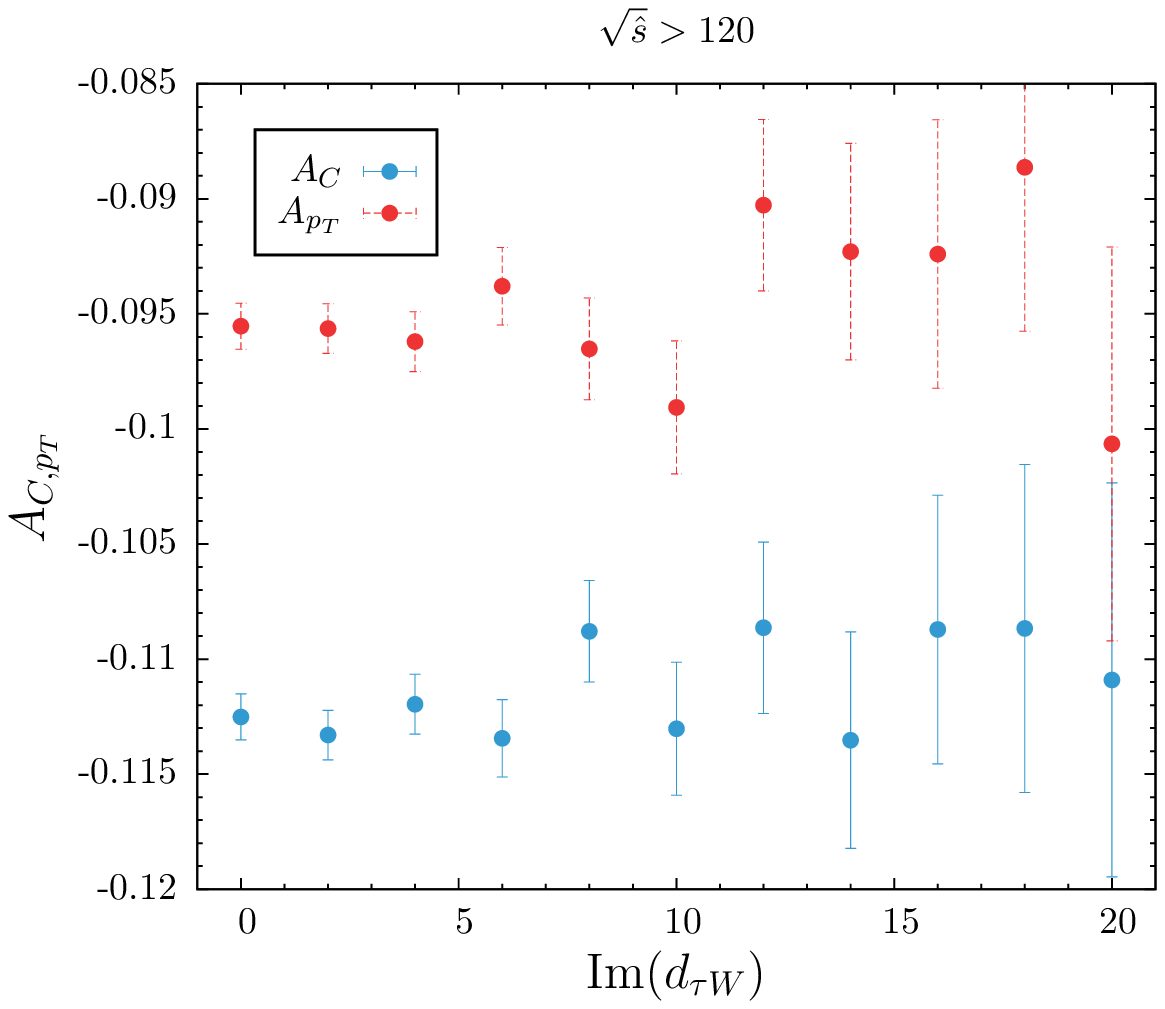} 
\caption{MC simulation of $A_{C,p_T}$ for high $m_{\tau\tau}$ events from Table~\ref{tiw} and also for events in the $Z$-region $60<m_{\tau\tau}<120$~GeV as a function of  ${\rm Im}(d_{\tau W})$.}
\label{fiw1011}
\end{figure}

\begin{table}[h]
\begin{center}
\begin{tabular}{|c|c|c|c|c|c|c|}\hline
Im($d_{\tau W}$) & $\sigma (\rm fb)$ & $A_1$ & $A_2$ & $A_{test}$ & $A_{C}$ & $A_{p_T}$ \\\hline
 0 & 95.45 &  0.0003 & 0.0000 & 0.0019  & -0.1125 & -0.0955\\\hline
 2 & 102.7 & -0.0283 & 0.0204 & -0.0002 & -0.1133 & -0.0956\\\hline
 4 & 124.3 & -0.0579 & 0.0399 & -0.0013 & -0.1119 & -0.0962\\\hline
 6 & 160.4 & -0.0858 & 0.0613 & -0.0018 & -0.1134 & -0.0938\\\hline
 8 & 210.9 & -0.1117 & 0.0780 & 0.0022  & -0.1088 & -0.0965\\\hline
10 & 276.0 & -0.1457 & 0.1019 & 0.0013  & -0.1131 & -0.0991\\\hline
12 & 355.4 & -0.1756 & 0.1131 & -0.0041 & -0.1086 & -0.0903\\\hline
14 & 449.2 & -0.1965 & 0.1334 & 0.0011  & -0.1135 & -0.0923\\\hline
16 & 557.0 & -0.2213 & 0.1693 & -0.0065 & -0.1087 & -0.0924\\\hline
18 & 680.0 & -0.2641 & 0.1745 & 0.0002  & -0.1087 & -0.0886\\\hline
20 & 817.0 & -0.2726 & 0.1793 & -0.0046 & -0.1109 & -0.1007\\\hline
\end{tabular}
\end{center}
\caption{Single spin $T$-odd correlations $A_{1,2}$ and $T$-even correlations $A_{C,p_T}$ for several values of ${\rm Im}(d_{\tau W})$ with ${\rm Re}(d_{\tau W})=0$. $A_{test}$ should vanish in all cases and gives us an estimate of the statistical error.}
\label{tiw}
\end{table}
In Table~\ref{tiw} we set ${\rm Re}(d_{\tau W})=0$ and fit the $T$-odd correlations to a linear equation. This is shown in Figure~\ref{fiw45} and the fits are consistent with the previous ones from Table~\ref{trwiw}. We tabulate $A_{test}$ again to assess the size of statistical fluctuations using an asymmetry that should be zero. Finally we also tabulate results for the $T$-even asymmetries which should not have a linear dependence on ${\rm Im}(d_{\tau W})$. This is confirmed in Figure~\ref{fiw1011} within our statistical uncertainty.

\begin{figure}[h]
\includegraphics[width=0.45\textwidth]{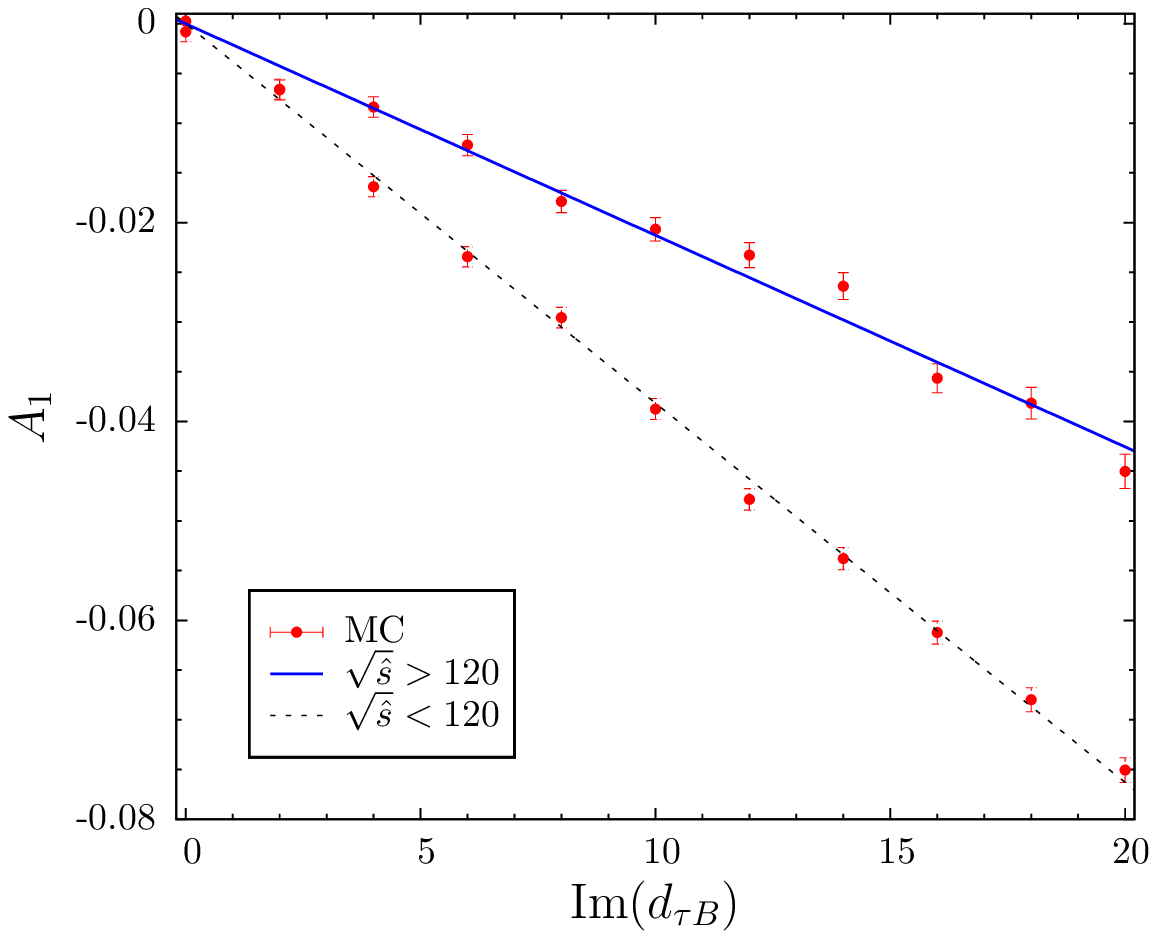} \hspace{0.5in}
\includegraphics[width=0.45\textwidth]{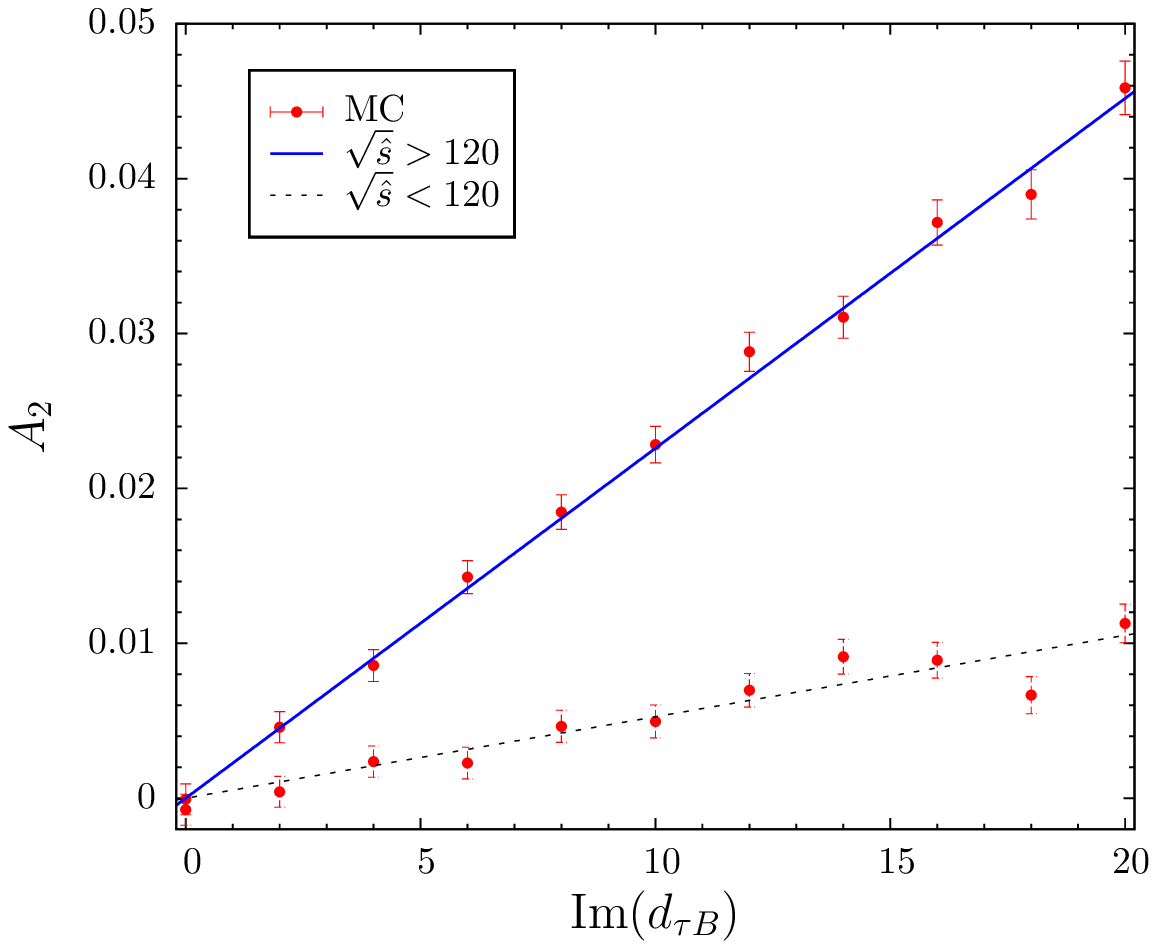}
\caption{MC simulation compared to a linear fit to  ${\rm Im}(d_{\tau B})$ for $A_1$ (left) and  $A_2$ (right)  for high $m_{\tau\tau}$ events from Table~\ref{tib} and also for events in the $Z$-region $60<m_{\tau\tau}<120$~GeV .}
\label{fib45}
\end{figure}

\begin{figure}[h]
\includegraphics[width=0.45\textwidth]{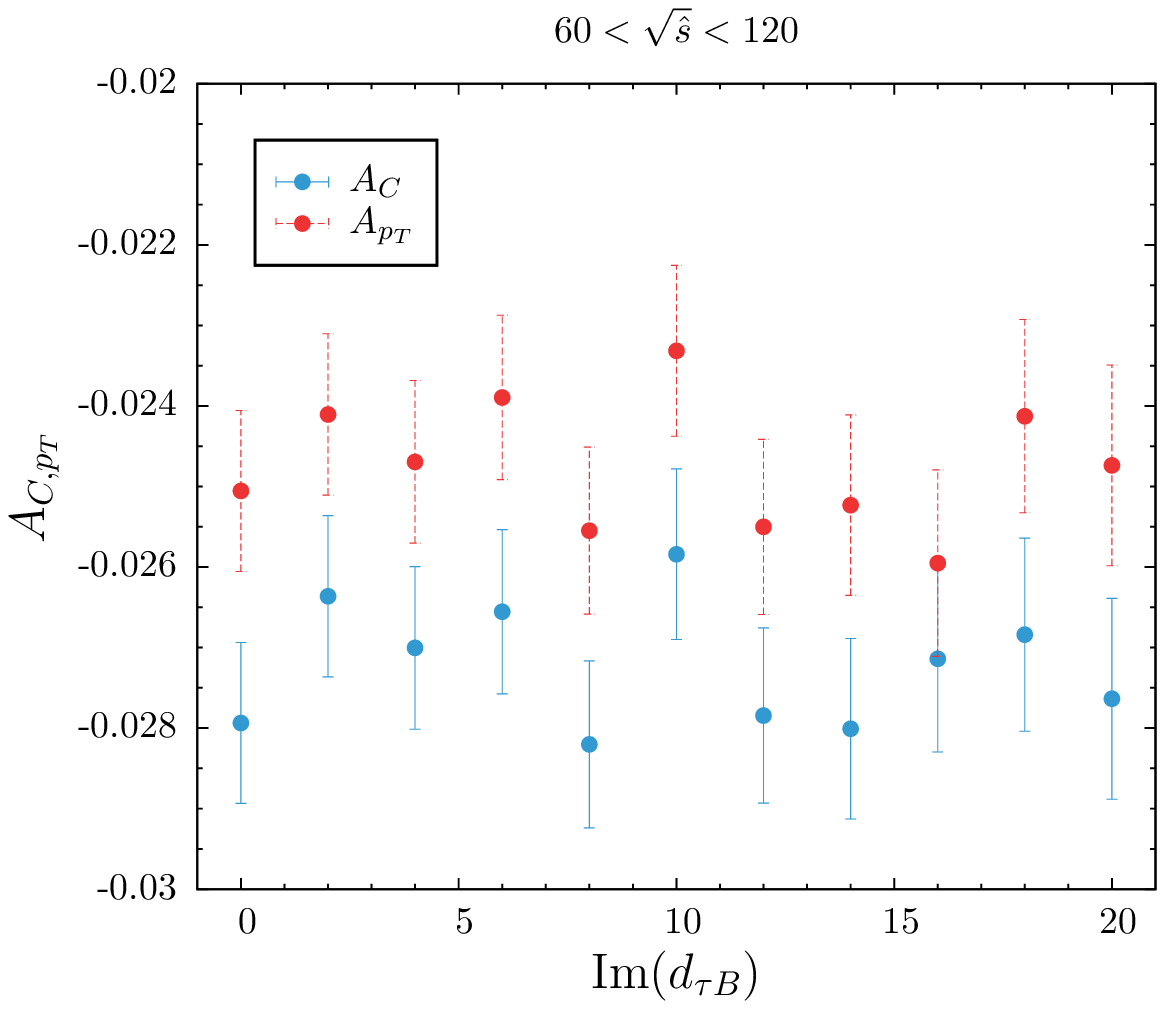} \hspace*{0.5cm}
\includegraphics[width=0.45\textwidth]{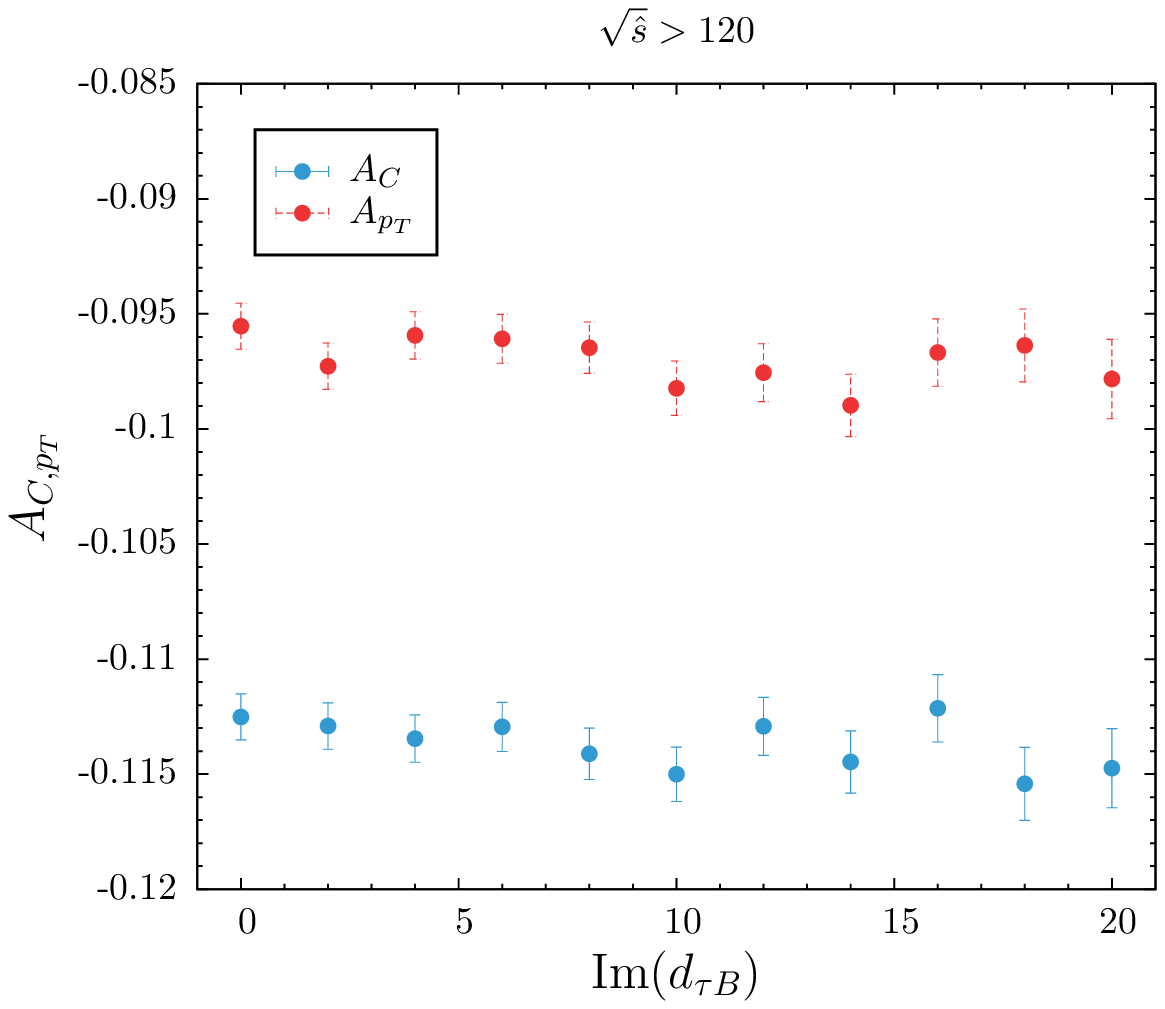}
\caption{MC simulation of $A_{C,p_T}$ for high $m_{\tau\tau}$ events from Table~\ref{tib} and also for events in the $Z$-region $60<m_{\tau\tau}<120$~GeV as a function of  ${\rm Im}(d_{\tau B})$.}
\label{fib1011}
\end{figure}

\begin{table}[h]
\begin{center}
\begin{tabular}{|c|c|c|c|c|c|c|}\hline
Im($d_{\tau B}$) & $\sigma (\rm fb)$ & $A_1$ & $A_2$ & $A_{test}$ & $A_{C}$ & $A_{p_T}$ \\\hline
 0 & 95.45 & 0.0003  & 0.0000 & 0.0019  & -0.1125 & -0.0955\\\hline
 2 & 96.16 & -0.0067 & 0.0046 & 0.0015  & -0.1129 & -0.0973\\\hline
 4 & 98.25 & -0.0083 & 0.0086 & 0.0018  & -0.1135 & -0.0959\\\hline
 6 & 101.7 & -0.0122 & 0.0143 & 0.0016  & -0.1129 & -0.0961\\\hline
 8 & 106.5 & -0.0179 & 0.0185 & 0.0015  & -0.1141 & -0.0964\\\hline
10 & 112.7 & -0.0207 & 0.0228 & 0.0007  & -0.1150 & -0.0982\\\hline
12 & 120.3 & -0.0233 & 0.0288 & 0.0012  & -0.1130 & -0.0976\\\hline
14 & 129.3 & -0.0264 & 0.0311 & -0.0008 & -0.1145 & -0.0990\\\hline
16 & 139.7 & -0.0357 & 0.0372 & -0.0013 & -0.1122 & -0.0967\\\hline
18 & 151.3 & -0.0382 & 0.0390 & -0.0013 & -0.1154 & -0.0964\\\hline
20 & 164.5 & -0.0450 & 0.0459 & 0.0007  & -0.1148 & -0.0979\\\hline
\end{tabular}
\end{center}
\caption{Single spin $T$-odd correlations $A_{1,2}$ and $T$-even correlations $A_{C,p_T}$ for several values of ${\rm Im}(d_{\tau B})$ with ${\rm Re}(d_{\tau B})=0$. $A_{test}$ should vanish in all cases and gives us an estimate of the statistical error.}
\label{tib}
\end{table}
Next we repeat the previous exercise but for ${\rm Im}(d_{\tau B})$ instead. The $T$-odd asymmetries are also linear in this coupling as expected, but smaller than those induced by ${\rm Im}(d_{\tau W})$ as can be seen in Figure~\ref{fib45}. Figure~\ref{fib1011} illustrates that the $T$-even asymmetries are not affected by this coupling within our numerical sensitivity.

\begin{figure}[h]
\includegraphics[width=0.45\textwidth]{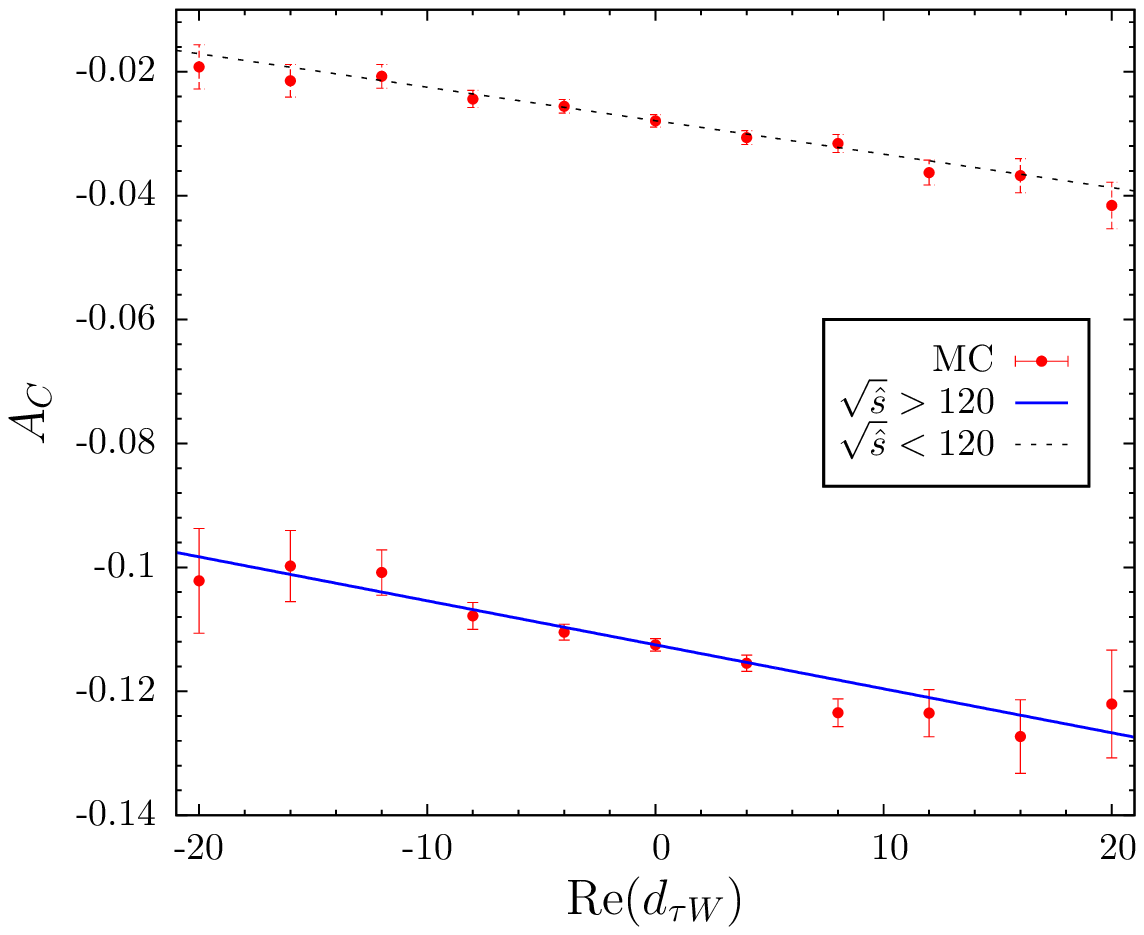} \hspace{0.5in}
\includegraphics[width=0.45\textwidth]{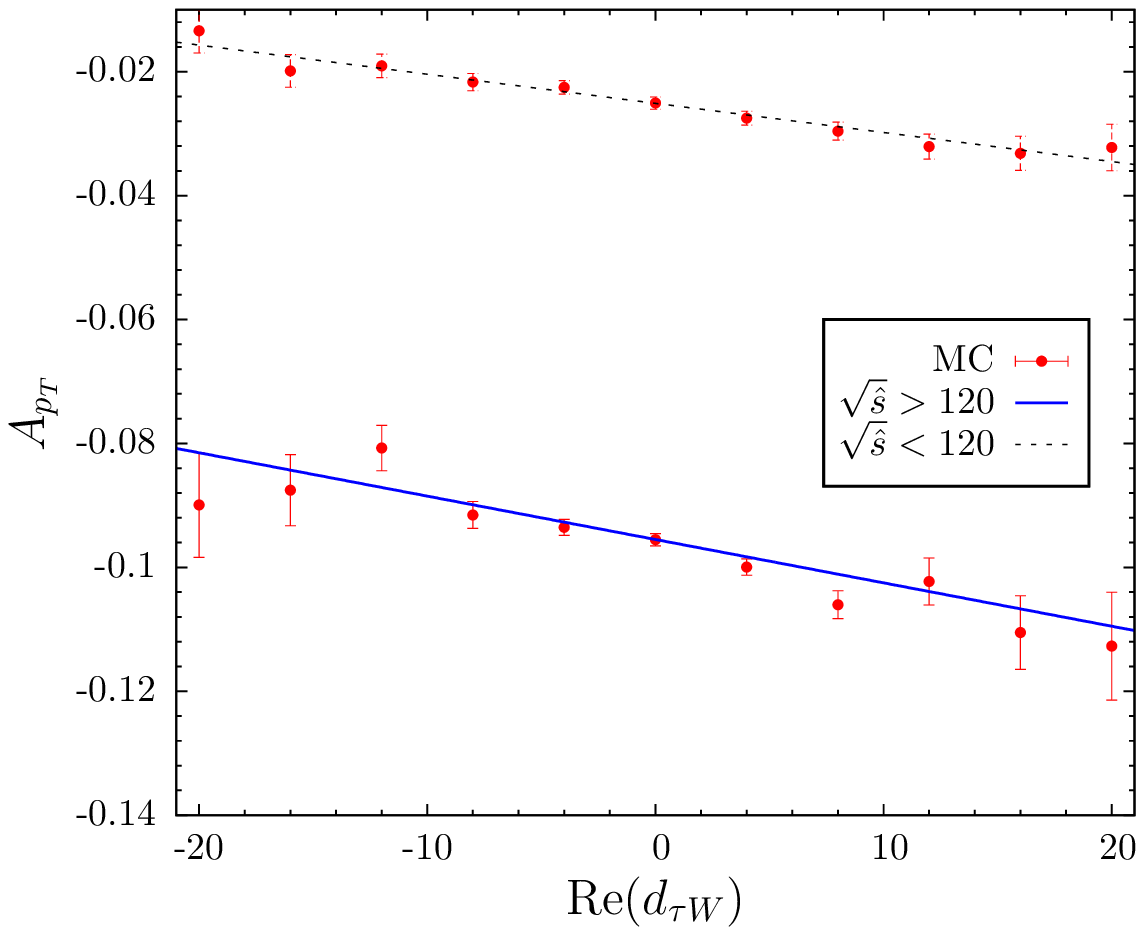}
\caption{MC simulation compared to a linear fit to  ${\rm Re}(d_{\tau W})$ for $A_{C}$ (left) and  $A_{p_T}$ (right) for high $m_{\tau\tau}$ events from Table~\ref{trw} and also for events in the $Z$-region $60<m_{\tau\tau}<120$~GeV.}
\label{firw45}
\end{figure}

\begin{figure}[h]
\includegraphics[width=0.45\textwidth]{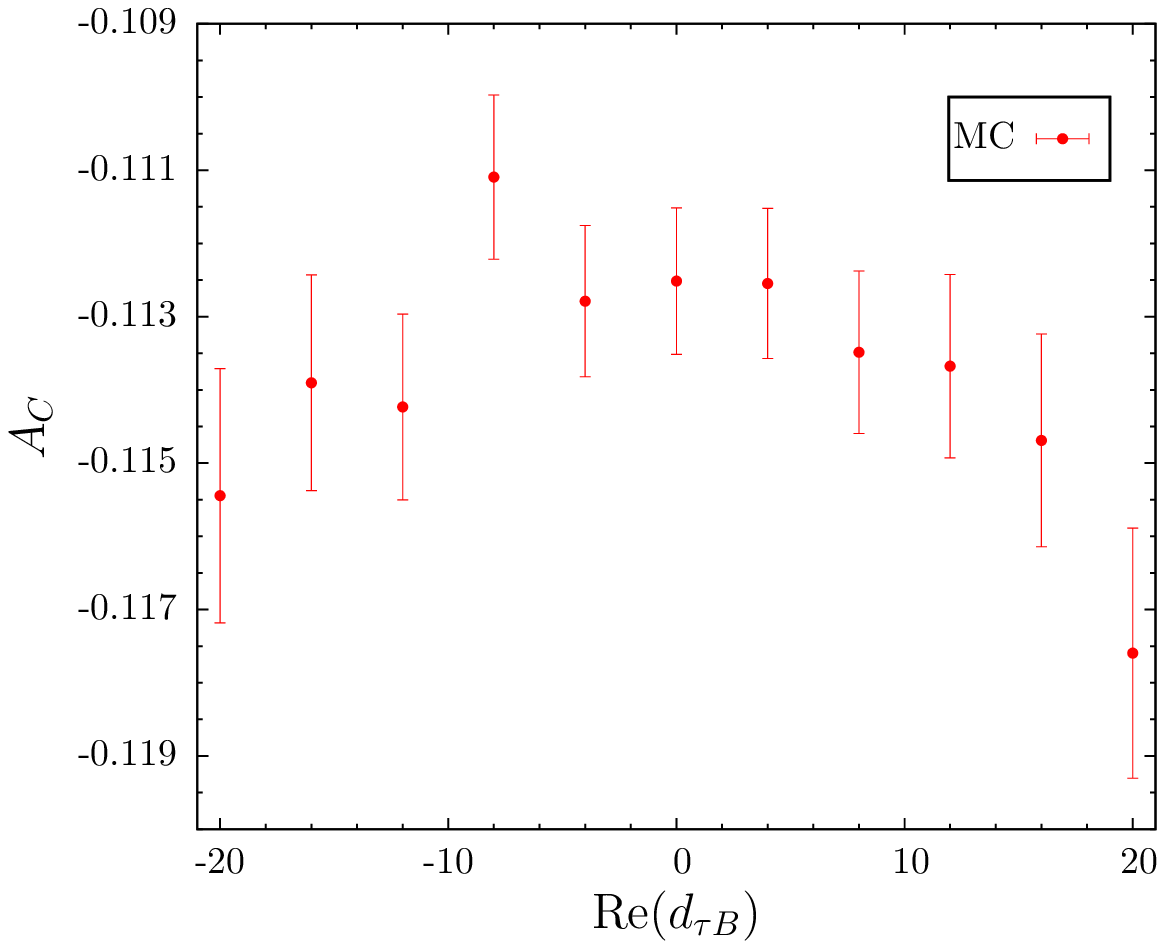} \hspace{0.5in}
\includegraphics[width=0.45\textwidth]{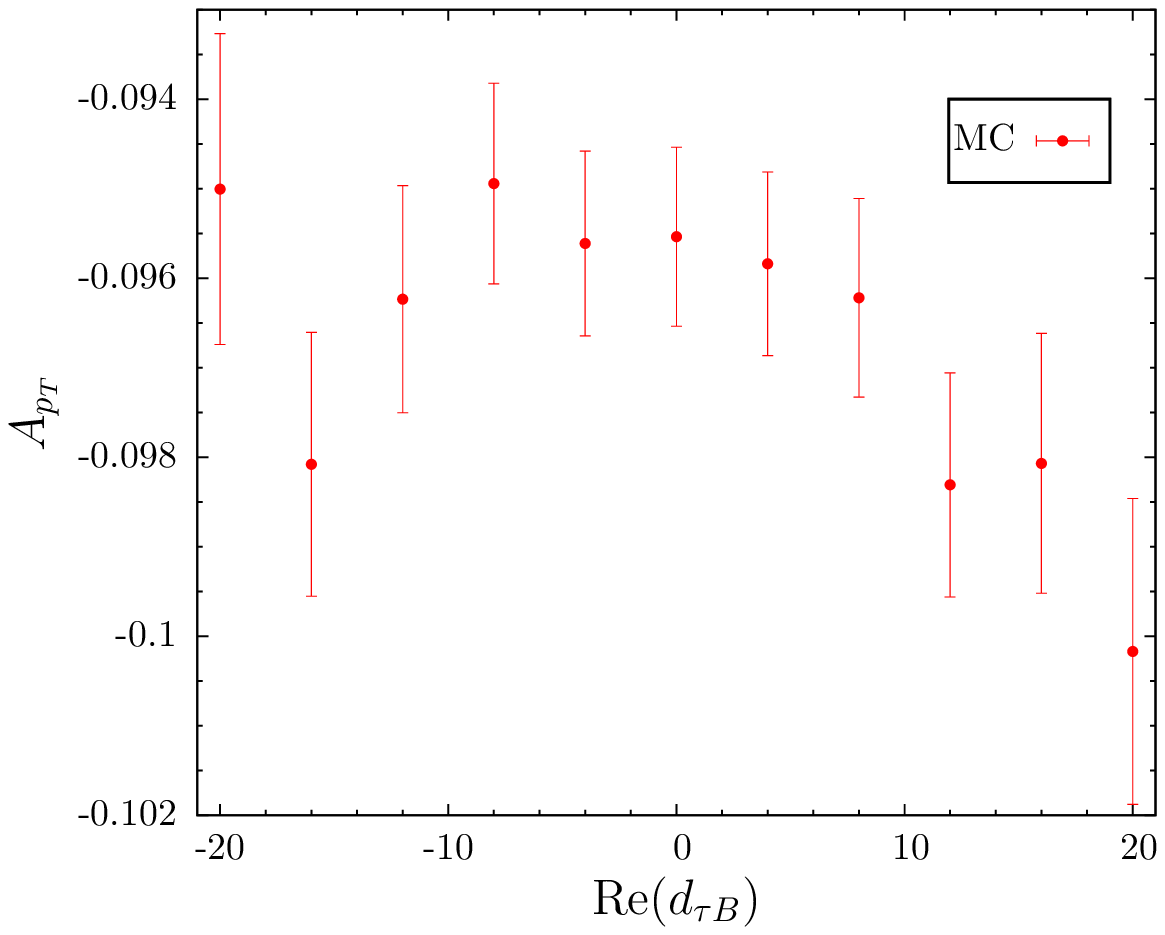}
\caption{MC simulation compared to a quadratic fit to  ${\rm Re}(d_{\tau B})$ for $A_{C}$ (left) and  $A_{p_T}$ (right) in Table~\ref{trb}.}
\label{firb1011}
\end{figure}

\begin{table}[h]
\begin{center}
\begin{tabular}{|c|c|c|c|c|c|c|}\hline
Re($d_{\tau W}$) & $\sigma (\rm fb)$ & $A_1$ & $A_2$ & $A_{test}$ & $A_{C}$ & $A_{p_T}$ \\\hline
-20 & 806.2 & -0.0059 & -0.0033 & -0.0208 & -0.1022 & -0.0899\\\hline
-16 & 548.8 & -0.0045 &  0.0007 &  0.0034 & -0.0998 & -0.0876\\\hline
-12 & 349.3 &  0.0000 & -0.0017 & -0.0034 & -0.1008 & -0.0807\\\hline
 -8 & 206.9 &  0.0015 &  0.0013 &  0.0004 & -0.1078 & -0.0915\\\hline
 -4 & 122.3 &  0.0017 & -0.0001 &  0.0002 & -0.1105 & -0.0935\\\hline
  0 & 95.45 &  0.0003 & -0.0000 &  0.0019 & -0.1125 & -0.0955\\\hline
  4 & 126.4 & -0.0026 &  0.0011 & -0.0007 & -0.1155 & -0.0999\\\hline
  8 & 215.1 & -0.0020 & -0.0005 &  0.0001 & -0.1235 & -0.1060\\\hline
 12 & 362.1 &  0.0045 &  0.0000 &  0.0014 & -0.1236 & -0.1023\\\hline
 16 & 566.9 &  0.0039 &  0.0036 &  0.0017 & -0.1273 & -0.1105\\\hline
 20 & 830.3 &  0.0145 &  0.0079 &  0.0019 & -0.1221 & -0.1127\\\hline
\end{tabular}
\end{center}
\caption{Single spin $T$-odd correlations $A_{1,2}$ and $T$-even correlations $A_{C,p_T}$ for several values of ${\rm Re}(d_{\tau W})$ with ${\rm Im}(d_{\tau W})=0$. $A_{test}$ should vanish in all cases and gives us an estimate of the statistical error.}
\label{trw}
\end{table}
We turn our attention to the $CP$ conserving couplings in Tables~\ref{trw} and \ref{trb}. We also find the expected behavior here: Figure~\ref{firw45} shows that the $T$-even asymmetries are linear in ${\rm Re}(d_{\tau W})$, and Table~\ref{trw}  shows that the $T$-odd asymmetries are all consistent with zero. The dependence of the $T$-even asymmetries on ${\rm Re}(d_{\tau B})$ is not observed within our statistical sensitivity as shown in Figure~\ref{firb1011}.

\begin{table}[h]
\begin{center}
\begin{tabular}{|c|c|c|c|c|c|c|}\hline
Re($d_{\tau B}$) & $\sigma (\rm fb)$ & $A_1$ & $A_2$ & $A_{test}$ & $A_{C}$ & $A_{p_T}$ \\\hline
-20 & 165.9 & -0.0002 & -0.0007 &  0.0016 & -0.1155 & -0.0950\\\hline
-16 & 140.7 & -0.0007 & -0.0028 & -0.0006 & -0.1139 & -0.0981\\\hline
-12 & 121.1 &  0.0001 &  0.0003 & -0.0017 & -0.1142 & -0.0962\\\hline
 -8 & 107.1 &  0.0008 & -0.0010 &  0.0005 & -0.1112 & -0.0950\\\hline
 -4 & 98.54 &  0.0005 &  0.0000 &  0.0005 & -0.1128 & -0.0956\\\hline
  0 & 95.45 &  0.0003 & -0.0000 &  0.0019 & -0.1125 & -0.0955\\\hline
  4 & 97.95 & -0.0000 & -0.0016 & -0.0008 & -0.1126 & -0.0958\\\hline
  8 & 105.9 & -0.0009 & -0.0000 &  0.0014 & -0.1135 & -0.0962\\\hline
 12 & 119.5 &  0.0003 &  0.0002 & -0.0002 & -0.1137 & -0.0983\\\hline
 16 & 138.6 &  0.0017 & -0.0004 &  0.0002 & -0.1147 & -0.0981\\\hline
 20 & 163.1 & -0.0035 & -0.0026 & -0.0026 & -0.1176 & -0.1002\\\hline
\end{tabular}
\end{center}
\caption{Single spin $T$-odd correlations $A_{1,2}$ and $T$-even correlations $A_{C,p_T}$ for several values of ${\rm Re}(d_{\tau B})$ with ${\rm Im}(d_{\tau B})=0$. $A_{test}$ should vanish in all cases and gives us an estimate of the statistical error.}
\label{trb}
\end{table}

\begin{table}[h]
\begin{center}
\begin{tabular}{|c|c|c|c|c|c|c|}\hline
$m_{\tau}$(GeV) & $\dfrac{\Gamma_\tau(m_\tau)}{\Gamma_\tau(2.5)}\cdot \left(\dfrac{2.5}{m_\tau}\right)^5$ & $Br(\tau^+\to 
\mu^+\nu_\mu\bar{\nu}_\tau)$ & $\sigma(\tau^+\tau^-)$ & $\sigma(\mu^+\mu^-\nu's)$ & $A_1$ & 
 $A_C$\\\hline
2.5 & - & 0.123 & 21.53 pb& 138.9 fb& -0.0721 &  -0.0589\\\hline
3.0 & 1.002 & 0.124 & 21.53 pb& 139.0 fb& -0.0696 &  -0.0561\\\hline
3.5 & 1.001 & 0.125 & 21.53 pb& 139.7 fb& -0.0709 &  -0.0576\\\hline
4.0 & 1.002 & 0.125 & 21.52 pb& 139.6 fb& -0.0729 &  -0.0549\\\hline
4.5 & 1.002 & 0.125 & 21.50 pb& 139.7 fb& -0.0725 &  -0.0585\\\hline
5.0 & 1.003 & 0.125 & 21.48 pb& 139.6 fb& -0.0697 & -0.0573\\\hline
5.5 & 1.004 & 0.125 & 21.46 pb& 139.5 fb& -0.0710 &  -0.0574\\\hline
6.0 & 1.005 & 0.125 & 21.45 pb& 139.4 fb& -0.0724 &  -0.0600\\\hline
6.5 & 1.006 & 0.125 & 21.43 pb& 139.2 fb& -0.0736 &  -0.0564\\\hline
7.0 & 1.007 & 0.125 & 21.40 pb& 139.2 fb& -0.0727 & -0.0583\\\hline
\end{tabular}
\end{center}
\caption{Scaling with the $\tau$ mass of cross-section and two asymmetries $A_1$ and $A_C$ for Im($d_{\tau W}$)=10. To remove any kinematic dependence from the decay vertices we use $m_u=m_d=m_c=m_s=0$, $m_b=10$ in all cases.} 
\label{taumass}
\end{table}
In Table~\ref{taumass} we study the dependence of some of the observables on the $\tau$-lepton mass. In order to keep kinematic factors in the $\tau$ decay constant, we also set the bottom-quark mass to be always higher than the $\tau$ mass and we take the charm, strange, up and down quarks as well as the muon to be massless. We see that the Drell-Yan cross section is approximately independent of the $\tau$-lepton mass, as expected. The width of the $\tau$-lepton exhibits the $m_\tau^5$ dependence predicted by the SM when the muon mass is neglected. The fact that $A_1$ is approximately constant in this table supports our interpretation of this result as originating mainly in the single spin asymmetry.

\begin{table}[h]
\begin{center}
\begin{tabular}{|c|c|c|c|c|c|}\hline
Collider & $\sigma (\rm fb)$ & $A_1$ & $A_2$ & $A_{test}$ & $A_{C}$ \\\hline
$pp$ & 276.0 & -0.1457 & 0.1019 & 0.0013  & -0.1131 \\\hline
$\bar{p}\bar{p}$ & 275.8 & -0.1418 & -0.0999 & -0.0021  & 0.1177 \\\hline
$p\bar{p}$ & 313.6 & -0.1531 & 0.0021 & 0.1687 & -0.0005 \\\hline
\end{tabular}
\end{center}
\caption{Comparison of $T$-odd and $T$-even asymmetries with Re($d_{\tau W}$)=0, Im($d_{\tau W}$)=10 for different colliders to exhibit their transformation properties under $CP$.}
\label{colliders}
\end{table}
In Table~\ref{colliders} we illustrate the $CP$ properties of the $T$-odd asymmetries discussed above. The asymmetry $A_1$ is $CP$-odd for the case of the $p\bar{p}$ collider whereas $A_2$ is $CP$-even and therefore cannot be induced by the anomalous coupling ${\rm Im}(d_{\tau W})$. For the LHC, a $pp$ collider, both $T$-odd asymmetries are possible as they transform into asymmetries in a $\bar{p}\bar{p}$ collider under a $CP$ transformation. They do so with opposite signs as can be seen in the Table. The charge asymmetry is $C$-odd and therefore changes sign at $\bar{p}\bar{p}$ collider and vanishes at a $p\bar{p}$ collider as seen in the example in Table~\ref{colliders}. $A_{test}$ on the other hand is not zero for $p\bar{p}$ colliders where the beam direction can be defined unambiguously.

\begin{table}[h]
\begin{center}
\begin{tabular}{|c|c|c|c|c|c|}\hline
Im($d_{\tau W}$) & $\sigma (\rm fb)$ & $A_1$ & $A_2$ & $A_{C}$ & $A_{p_T}$ \\\hline
10 & 1111.0 & -0.1363 & 0.0974 & -0.1100 & -0.0956 \\\hline
\end{tabular}
\end{center}
\caption{Selected asymmetries in the dilepton channel ($pp\to \tau^+ \tau^- \to \ell^+ \ell^- \nu{\rm 's}$) with Re($d_{\tau W}$)=0. In this case $\sigma_{\rm SM}=385.2$ fb.}
\label{muande}
\end{table}
Table~\ref{muande} illustrates that replacing the dimuon channel with the dilepton channel simply increases the statistics by a factor of four and does not change the four asymmetries we have been discussing.

\begin{table}[h]
\begin{center}
\begin{tabular}{|c|c|c|c|c|}\hline
$\Gamma_\tau$(GeV) & Im($d_{\tau W}$) & $\sigma (\rm fb)$ & $A_2$ & $A_{C}$ \\\hline
$2.27\times 10^{-5}$ & 10 & 276.0 & 0.1019 & -0.1131 \\\hline
$2.27\times 10^{-6}$ & 10 & 276.0 & 0.0952 & -0.1156 \\\hline
$2.27\times 10^{-7}$ & 10 & 275.8 & 0.0960 & -0.1179 \\\hline
$2.27\times 10^{-8}$ & 10 & 275.9 & 0.0953 & -0.1169 \\\hline
$2.27\times 10^{-9}$ & 10 & 276.0 & 0.0975 & -0.1155 \\\hline
\end{tabular}
\end{center}
\caption{Effect of changing the $\tau$-lepton width in the MC simulation. After the rescaling described in the text the cross-section as well as the asymmetries are seen to be independent of $\Gamma_\tau$ within our numerical accuracy. In all cases we took Re($d_{\tau W}$)=0.}
\label{tauwidth}
\end{table}
In Table~\ref{tauwidth} we demonstrate that the trick of using a fictitious $\tau$-lepton width in the simulations does not affect the cross-section or the asymmetries $A_2$ and $A_{C}$ (it also does not affect the other asymmetries).

\begin{figure}[h]
\includegraphics[width=0.45\textwidth]{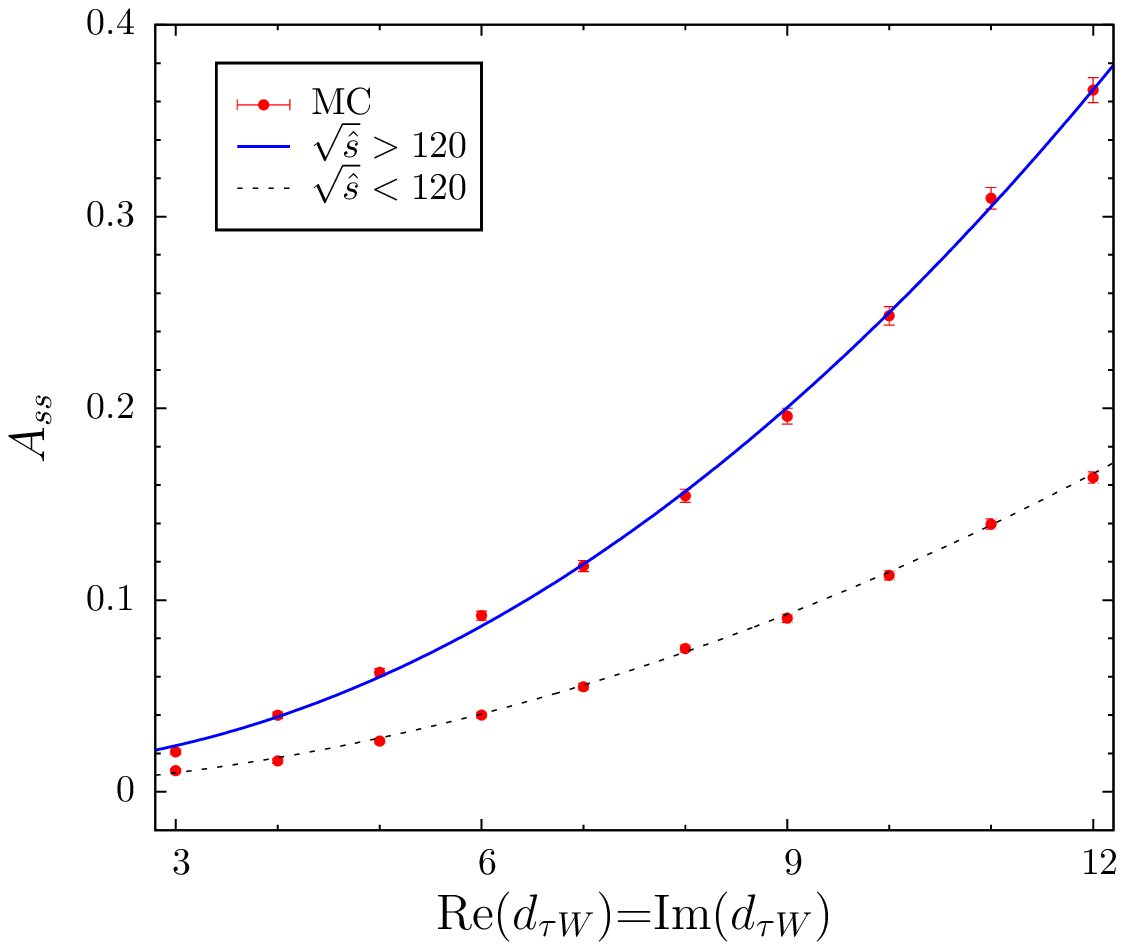} \\ \vspace{0.2in}
\includegraphics[width=0.45\textwidth]{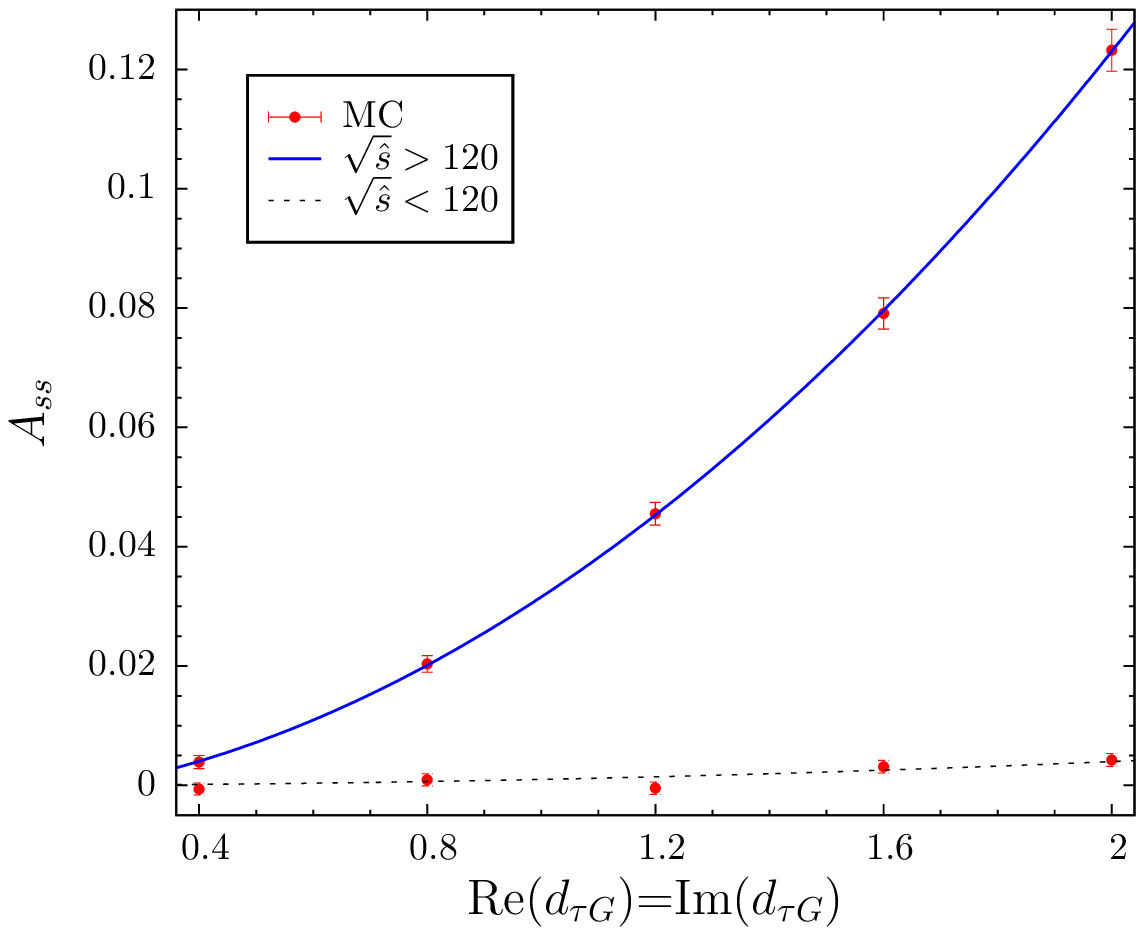}\hspace{0.5in}
\includegraphics[width=0.45\textwidth]{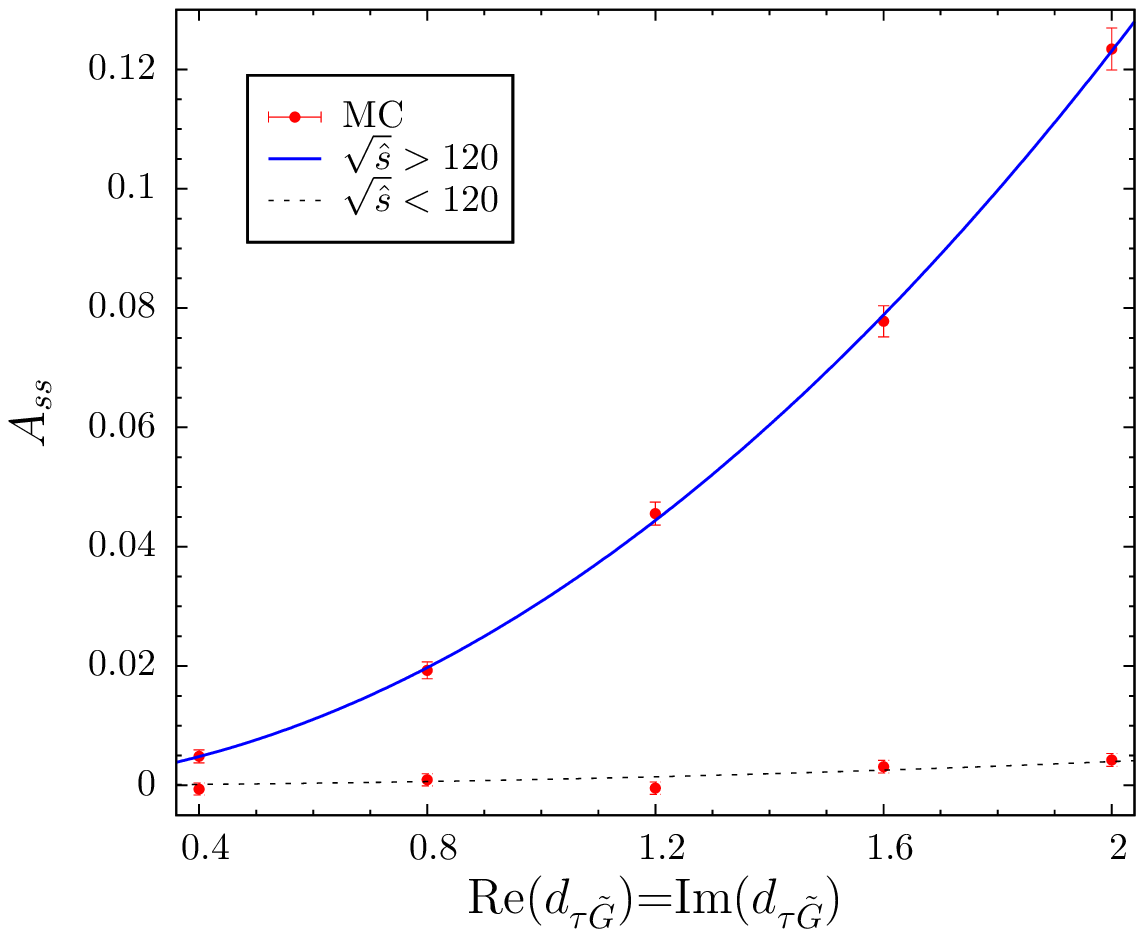}
\caption{Comparison of MC and a fit linear in  ${\rm Re}(d_{\tau W}){\rm Im}(d_{\tau W})$  (up), ${\rm Re}(d_{\tau G}){\rm Im}(d_{\tau G})$, (left) and ${\rm Re}(d_{\tau \tilde{G}}){\rm Im}(d_{\tau  \tilde{G}})$ (right), for $A_{ss}$ shown separately for events with high $m_{\tau\tau}$  and  for events in the $Z$-region $60<m_{\tau\tau}<120$~GeV .}
\label{fi2s}
\end{figure}

\begin{table}[h]
\begin{center}
\begin{tabular}{|c|c|c|c|}\hline
Re($d_{\tau W}$) & Im($d_{\tau W}$) & $\sigma (\rm fb)$ & $A_{ss}$ \\\hline
3 & 3 & 129.5 & 0.0209 \\\hline
4 & 4 & 155.3 & 0.0399 \\\hline
5 & 5 & 188.4 & 0.0623 \\\hline
6 & 6 & 228.6 & 0.0919 \\\hline
7 & 7 & 276.2 & 0.1177 \\\hline
8 & 8 & 331.0 & 0.1544 \\\hline
9 & 9 & 392.9 & 0.1959 \\\hline
10 & 10 & 462.2 & 0.2483 \\\hline
11 & 11 & 538.5 & 0.3095 \\\hline
12 & 12 & 622.2 & 0.3660 \\\hline
\end{tabular}
\end{center}
\caption{Double spin correlation $A_{ss}$ induced by interference between ${\rm Re}(d_{\tau W})$ and ${\rm Im}(d_{\tau W})$.}
\label{reimw}
\end{table}
In Table~\ref{reimw}, we set ${\rm Re}(d_{\tau W})={\rm Im}(d_{\tau W})$ to look for the double spin asymmetry through $A_{ss}$. The result of the fit is as shown in Figure~\ref{fi2s}.
\begin{table}[h]
\begin{center}
\begin{tabular}{|c|c|c|c|c|c|}\hline
Re($d_{\tau G}$) & Im($d_{\tau G}$) & $\sigma (\rm fb)$ & $A_{ss}$ & $A_{C}$ & $A_{p_T}$ \\\hline
0.4 & 0.4 & 105.1 &  0.0039 & -0.1123 & -0.0955 \\\hline
0.8 & 0.8 & 134.1 &  0.0204 & -0.1152 & -0.0975 \\\hline
1.2 & 1.2 & 182.2 & -0.0455 & -0.1126 & -0.0966 \\\hline
1.6 & 1.6 & 249.8 &  0.0791 & -0.1118 & -0.0948 \\\hline
2.0 & 2.0 & 336.7 &  0.1232 & -0.1084 & -0.0923 \\\hline
\end{tabular}
\end{center}
\caption{Double spin correlation $A_{ss}$ induced by interference between ${\rm Re}(d_{\tau G})$ and ${\rm Im}(d_{\tau G})$. No discernible effect from these couplings is found in other asymmetries.}
\label{reimg}
\end{table}
In Table~\ref{reimg}, we set ${\rm Re}(d_{\tau G})={\rm Im}(d_{\tau G})$ to look for the double spin asymmetry through $A_{ss}$. The result of the fit is as shown in Figure~\ref{fi2s}.

\begin{table}[h]
\begin{center}
\begin{tabular}{|c|c|c|c|c|c|}\hline
Re($d_{\tau \tilde{G}}$) & Im($d_{\tau \tilde{G}}$) & $\sigma (\rm fb)$ & $A_{ss}$ & $A_{C}$ & $A_{p_T}$ \\\hline
0.4 & 0.4 & 105.1 & 0.0049 & -0.1132 & -0.0945 \\\hline
0.8 & 0.8 & 134.1 & 0.0193 & -0.1131 & -0.0964 \\\hline
1.2 & 1.2 & 182.2 & 0.0455 & -0.1124 & -0.0974 \\\hline
1.6 & 1.6 & 249.8 & 0.0778 & -0.1155 & -0.0971 \\\hline
2.0 & 2.0 & 336.7 & 0.1234 & -0.1117 & -0.0951 \\\hline
\end{tabular}
\end{center}
\caption{Double spin correlation $A_{ss}$ induced by interference between ${\rm Re}(d_{\tau \tilde{G}})$ and ${\rm Im}(d_{\tau \tilde{G}})$. No discernible effect from these couplings is found in other asymmetries.}
\label{reimgg}
\end{table}
In Table~\ref{reimgg}, we set ${\rm Re}(d_{\tau \tilde{G}})={\rm Im}(d_{\tau \tilde{G}})$ to look for the double spin asymmetry through $A_{ss}$. The result of the fit is as shown in Figure~\ref{fi2s}.

\begin{table}[h]
\begin{center}
\begin{tabular}{|c|c|c|c|c|c|}\hline
Re($d_{\tau G}$) & Im($d_{\tau \tilde{G}}$) & $\sigma (\rm fb)$ & $A_{ss}$ & $A_{C}$ & $A_{p_T}$ \\\hline
2.0 & 2.0 & 336.9 & -0.0058 & -0.1116 & -0.0956 \\\hline\hline
Re($d_{\tau \tilde{G}}$) & Im($d_{\tau G}$) & $\sigma (\rm fb)$ & $A_{ss}$ & $A_{C}$ & $A_{p_T}$ \\\hline
2.0 & 2.0 & 336.9 & -0.0026 & -0.1151 & -0.0987 \\\hline
\end{tabular}
\end{center}
\caption{The real and imaginary parts of the different couplings $d_{\tau G}$ and $d_{\tau \tilde{G}}$ do not show interference effects.}
\end{table}

\end{document}